\documentclass[final,5p,times,twocolumn]{elsarticle}

\usepackage{amssymb}
\usepackage{amsmath}
\usepackage{lineno}
\usepackage{url}
\usepackage{hyperref}
\usepackage{multirow}
\usepackage{booktabs}
\usepackage{makecell}
\usepackage{pifont} 
\usepackage{ulem}
\usepackage{algorithm}
\usepackage{algorithmic}
\usepackage{amsmath} 

\journal{xxx}

\begin{document}

\begin{frontmatter}



\title{EndoControlMag: Robust Endoscopic Vascular Motion Magnification with Periodic Reference Resetting and Hierarchical Tissue-aware Dual-Mask Control\tnoteref{fund}}

\tnotetext[fund]{This work was supported by Hong Kong Research Grants Council (RGC) Collaborative Research Fund (C4026-21G), General Research Fund (GRF 14211420 \& 14203323), Shenzhen-Hong Kong-Macau Technology Research Programme (Type C) STIC Grant SGDX20210823103535014 (202108233000303).}

\author[1,2]{An Wang\fnref{equal}} 
\author[2]{Rulin Zhou\fnref{equal}}
\author[1]{Mengya Xu\fnref{equal}}
\author[3]{Yiru Ye}
\author[4]{Longfei Gou}
\author[1]{Yiting Chang}
\author[4]{Hao Chen}
\author[5]{Chwee Ming Lim}
\author[6]{\\Jiankun Wang}
\author[1,2]{Hongliang Ren\corref{cor1}}

\fntext[equal]{Equal Contribution.}
\cortext[cor1]{Corresponding Author. \ead{hlren@ee.cuhk.edu.hk.}}
\affiliation[1]{organization={Department of Electronic Engineering, The Chinese University of Hong Kong},
            state={Hong Kong SAR},
            country={China}}
\affiliation[2]{organization={The Chinese University of Hong Kong Shenzhen Research Institute},
            city={Shen Zhen},
            state={Guangdong},
            country={China}}
\affiliation[3]{organization={The First Affiliated Hospital of Wenzhou Medical University},
            city={Wenzhou},
            state={Zhejiang},
            country={China}} 
\affiliation[4]{organization={Department of General Surgery \& Guangdong Provincial Key Laboratory of Precision Medicine for Gastrointestinal Tumor, Nanfang Hospital, Southern Medical University},
            city={Guang Zhou},
            state={Guangdong},
            country={China}}
\affiliation[5]{organization={Department of Otolaryngology-Head and Neck Surgery, Singapore General Hospital, Duke-NUS Medical School},
            country={Singapore}}
\affiliation[6]{organization={Shenzhen Key Laboratory of Robotics Perception and Intelligence, and the Department of Electronic and Electrical Engineering, Southern University of Science and Technology},
            city={Shenzhen},
            country={China}}
            

\begin{abstract}
Visualizing subtle vascular motions in endoscopic surgery is crucial for surgical precision and decision-making, yet remains challenging due to the complex and dynamic nature of surgical scenes. To address this, we introduce \textbf{EndoControlMag}, a training-free, Lagrangian-based framework with mask-conditioned vascular motion magnification tailored to endoscopic environments. Our approach features two key modules: a \textit{Periodic Reference Resetting (PRR)} scheme that divides videos into short overlapping clips with dynamically updated reference frames to prevent error accumulation while maintaining temporal coherence, and a \textit{Hierarchical Tissue-aware Magnification (HTM)} framework with dual-mode mask dilation. HTM first tracks vessel cores using a pretrained visual tracking model to maintain accurate localization despite occlusions and view changes. It then applies one of two adaptive softening strategies to surrounding tissues: motion-based softening that modulates magnification strength proportional to observed tissue displacement, or distance-based exponential decay that simulates biomechanical force attenuation. This dual-mode approach accommodates diverse surgical scenarios—motion-based softening excels with complex tissue deformations while distance-based softening provides stability during unreliable optical flow conditions.
We evaluate EndoControlMag on our \textbf{EndoVMM24} dataset spanning four different surgery types and various challenging scenarios, including occlusions, instrument disturbance, view changes, and vessel deformations. 
Quantitative metrics, visual assessments, and expert surgeon evaluations demonstrate that EndoControlMag significantly outperforms existing methods in both magnification accuracy and visual quality while maintaining robustness across challenging surgical conditions. The code, dataset, and video results are available at~\url{https://szupc.github.io/EndoControlMag/}.
\end{abstract}

\begin{keyword}
Vascular Motion Magnification \sep 
Endoscopic Vision Enhancement \sep 
Conditioned Video Editing \sep 
Periodic Reference Resetting \sep 
Hierarchical Mask Dilation

\end{keyword}

\end{frontmatter}

\section{Introduction}

\begin{figure}[ht]
    \centering
    \includegraphics[width=0.8\linewidth]{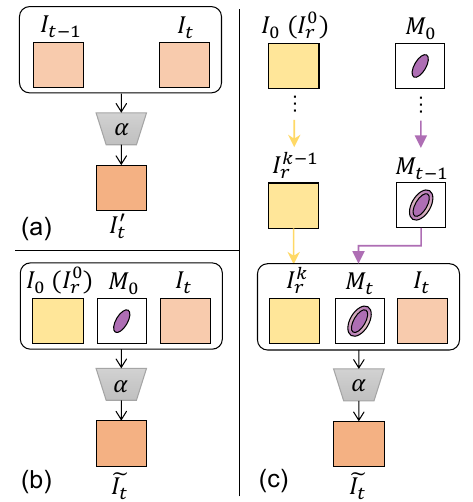}
    \label{fig:controlmag}
    \caption{\textbf{Comparison of motion magnification approaches for transforming an input frame $I_t$ into a magnified frame $\tilde{I_t}$ with a factor $\alpha$.} (a) Conventional methods apply global magnification uniformly to the entire image without region-specific control. (b) FlowMag~\cite{pan2024self} introduces mask-conditioned magnification but relies on a fixed reference frame $I_0$ and static mask $M_0$ throughout the sequence, leading to mask misalignment and abrupt boundary transitions. (c) Our EndoControlMag addresses these limitations through periodically resetting the reference frame and employing a dual-mask strategy, which recursively tracks the inner mask while applying softened dilation to the outer mask.}
\end{figure}

Endoscopic surgery has transformed the field of minimally invasive procedures, offering enhanced precision and reduced patient recovery times~\cite{mack2001minimally}. However, one of the persistent challenges in endoscopic surgery is the accurate visualization of subtle vascular dynamics, which play a crucial role in surgical decision-making~\cite{janatka2018higher,yang2024magnification,janatka2020surgical}. Surgeons often rely on their expertise to interpret these subtle cues from live video feeds, but human eyes can struggle to detect minute vascular pulsations amidst the complex and dynamic surgical environment~\cite{shander2007financial}, where various challenges~\cite{ross2020robust,ding2024segstrong} like electrocautery-induced smoke, instrument occlusions, and huge view shifts complicate the clear visualization of vascular motion.
This limitation necessitates advanced visualization technologies capable of amplifying these critical signals without introducing distracting artifacts~\cite{janatka2018higher,janatka2020surgical,fan2021robotically,huang2023motion}.

To address this challenge, video motion magnification (VMM)~\cite{liu2005motion,wu2012eulerian,janatka2018higher} techniques have emerged as a promising solution. These methods amplify subtle motions in video sequences, making imperceptible movements visible to the naked eye. 
Traditional methods for enhancing motion visualization in videos include Eulerian approaches~\cite{wu2012eulerian,ahmed2023overview,wang2024eulermormer}, which amplify temporal intensity variations at fixed pixel locations, and Lagrangian approaches~\cite{liu2005motion,flotho2023lagrangian}, which explicitly estimate and magnify motion paths. While recent deep learning VMM approaches~\cite{oh2018learning,ha2024revisiting,lado2023stb,byung2025learning,pan2024self} show potential, they typically demand extensive, domain-specific training data, which is scarce in surgery, and often lack the adaptability and interactive control needed for real-time surgical guidance. Crucially, most existing methods apply magnification globally, failing to provide the targeted enhancement required for specific vascular structures within a dynamic surgical scene.

In this work, we introduce \textbf{EndoControlMag}, a mask-conditioned, training-free Lagrangian motion magnification framework specifically designed to enhance vascular pulsation visibility in endoscopic surgery.
As illustrated in Fig.~\ref{fig:controlmag}, compared with conventional approaches which apply global uniform magnification across the entire image and the former baseline~\cite{pan2024self} which relies on a fixed reference frame and static mask, our EndoControlMag advances beyond these limitations through two key designs: \textit{Periodic Reference Resetting (PRR)} and \textit{Hierarchical Tissue-aware Magnification (HTM)}. Specifically, the PRR scheme divides video sequences into short overlapping clips, with the first frame of each clip serving as an updated reference. This approach prevents error accumulation in motion estimation while maintaining temporal coherence throughout the surgical procedure. Besides, the HTM framework incorporates video object tracking to dynamically follow vessel movements, ensuring consistent magnification despite camera motions, tissue manipulation, and occlusions. This tracking mechanism is complemented by a dual-mask strategy that distinguishes between the core vascular structure and surrounding tissue. We implement two adaptive softening approaches for the outer region: a motion-based strategy that modulates magnification strength proportionally to observed tissue displacement, and a distance-based strategy that implements exponential decay from vessel boundaries to simulate biomechanical force attenuation. This design ensures smooth transitions between magnified and non-magnified areas, minimizing artifacts while respecting the deformable nature of biological tissues.

To provide a comprehensive assessment of the magnification performance, we construct the \textbf{EndoVMM24} dataset, spanning four different surgical specialties and various challenging scenarios including instrument occlusions, view changes, vessel deformations, and tool disturbance. Both quantitative metrics and expert surgical assessments demonstrate that EndoControlMag significantly outperforms existing methods in magnification accuracy, visual quality, and robustness across diverse surgical conditions.

In conclusion, this work advances surgical vision enhancement through a controllable, robust, and context-aware vascular motion magnification framework. The key contributions are:
\begin{itemize}
    \item We present \textbf{EndoControlMag}, a training-free Lagrangian vascular motion magnification framework incorporating mask-conditioned control for enhanced visibility and interactivity in endoscopic surgery.
    \item We introduce \textit{Periodic Reference Resetting (PRR)} and \textit{Hierarchical Tissue-aware Magnification (HTM)}, integrating error-resettable reference, dynamic mask tracking, and adaptive softening strategies for robust, context-aware artifact minimization specifically tailored to the complex demands of surgical environments.
    \item We construct \textbf{EndoVMM24}, a comprehensive dataset encompassing multiple surgical specialties and challenging clinical scenarios. Through extensive quantitative evaluation and expert surgical assessment, we demonstrate the effectiveness, robustness, and potential clinical applicability of EndoControlMag across diverse surgical procedures.
\end{itemize}

By addressing the unique challenges of endoscopic surgical visualization, EndoControlMag offers a promising approach to enhance surgical vision during minimally invasive procedures, potentially contributing to improved visual feedback during critical phases of operation.

\section{Related Works}

\subsection{Video Motion Magnification}
Video Motion Magnification (VMM) aims to amplify subtle motions in video sequences, rendering imperceptible movements visible~\cite{liu2005motion}. 
Traditional approaches can be broadly classified into Eulerian~\cite{wu2012eulerian,ahmed2023overview,wang2024eulermormer} and Lagrangian~\cite{liu2005motion,flotho2023lagrangian} methods. 
Eulerian methods operate on intensity variations at fixed pixel locations over time. Wu et al.~\cite{wu2012eulerian} introduced Eulerian Video Magnification (EVM), employing a Laplacian pyramid to decompose frames and amplify temporal intensity variations. While computationally efficient for subtle motions, EVM introduces significant noise and artifacts under surgical conditions with dynamic lighting and tissue deformation. Wadhwa et al.~\cite{wadhwa2013phase} proposed a phase-based method using complex steerable pyramids to improve sensitivity and reduce noise. However, phase-based approaches still struggle with larger motions and generate ringing artifacts, limiting their applicability in dynamic surgical environments. Subsequent Eulerian advancements have focused on improving noise reduction~\cite{janatka2020surgical} and enhancing motion representation~\cite{janatka2018higher,zhang2017video}, but remain fundamentally limited by their inability to handle large displacements.

Lagrangian methods track pixels or regions explicitly over time, modeling motion trajectories with greater robustness to large motions and occlusions~\cite{liu2005motion,flotho2023lagrangian}. Liu et al.~\cite{liu2005motion} pioneered this approach by clustering pixels based on motion similarity and amplifying trajectories. However, their method requires manual intervention and significant computational resources, limiting clinical applicability. Recent efforts have focused on automating Lagrangian methods through optical flow estimation and deep learning. Fan et al.~\cite{fan2021robotically} presented a hybrid approach combining temporal filtering with deep spatial decomposition to enhance vascular pulsations while reducing noise. Pan et al.~\cite{pan2024self} introduced a self-supervised framework leveraging optical flow networks for motion amplification without requiring labeled data.

While traditional handcrafted filter methods~\cite{zhang2017video,takeda2022bilateral} often necessitate extensive hyperparameter tuning to achieve optimal performance, the integration of deep learning has further enhanced motion magnification techniques~\cite{oh2018learning,ha2024revisiting,lado2023stb,pan2024self,byung2025learning,gao2022magformer,wang2024eulermormer}, enabling end-to-end learning and hierarchical feature extraction. For instance, DMM~\cite{oh2018learning} developed a learning-based model using synthetic data, effectively reducing noise but facing generalization challenges in diverse surgical scenarios. Recent advancements, such as STB-VMM~\cite{lado2023stb} and Axial-VMM~\cite{byung2025learning} have utilized Swin Transformers~\cite{liu2021swin} and attention mechanisms~\cite{vaswani2017attention} to improve feature learning and magnification quality. Concurrently, hybrid architectures like MagFormer~\cite{gao2022magformer} synergize Eulerian and Lagrangian paradigms, demonstrating that complementary frameworks can yield enhanced performance by merging their respective advantages.

Despite these advancements, most existing VMM methods apply magnification globally, failing to provide the targeted enhancement required for specific vascular structures within dynamic surgical scenes. Additionally, deep learning-based methods face limitations due to their substantial data requirements, compounded by the scarcity of annotated surgical datasets—a key barrier to implementing data-driven solutions in this domain.

\subsection{Conditioned Video Editing}
Conditioned video editing has emerged as a powerful paradigm for manipulating or synthesizing videos based on specific input signals, offering finer control compared to traditional processing techniques. Various forms of signals, including texts, segmentation masks, depth maps, and optical flow, can be leveraged as the control conditions~\cite{zhang2023adding,wu2023tune,chen2023control,wang2024easycontrol,yan2023motion,duan2024diffutoon,zheng2024makima,geng2024motion}. Among these, mask-conditioned editing has gained particular traction in computer vision, enabling precise spatial control over video manipulation. Applications include video object inpainting~\cite{zhou2023propainter,zhang2024avid,daher2023temporal}, where masks guide the removal and replacement of content; video style transfer~\cite{song2024univst,duan2024diffutoon}, where masks designate regions for stylistic modification; and video harmonization~\cite{lu2022deep,xiao2024tsa2}, where masks define areas requiring seamless integration of composited elements.

In the context of motion magnification, FlowMag~\cite{pan2024self} recently pioneered mask-based conditioning for targeted video motion magnification, allowing selective amplification of motions within user-defined regions. However, this approach is constrained by its reliance on a static mask that remains fixed throughout the video sequence. This limitation is particularly problematic in dynamic surgical environments, where camera movements and tissue manipulation cause constant spatial reconfiguration. Additionally, FlowMag does not account for the biomechanical relationship between vascular structures and surrounding tissues, treating the boundary between magnified and unmagnified regions as a binary transition rather than modeling the graduated influence of vascular pulsations on adjacent tissues.
These limitations underscore the need for a more sophisticated approach to mask-conditioned motion magnification in surgical contexts—one that can dynamically track regions of interest while modeling the complex biomechanical interactions between vascular structures and surrounding tissues.

\subsection{Limitations of Existing Methods and Our Contribution}
Existing motion magnification methods face significant limitations when applied to endoscopic surgery videos. Traditional approaches rely on manually tuned filters with fixed parameters that cannot adapt to the dynamic and heterogeneous nature of surgical scenes. Learning-based methods require extensive training data and struggle to generalize across diverse surgical procedures and anatomical structures. Both approaches typically operate globally or with static masks, amplifying motion uniformly across the frame without considering the biomechanical relationships between tissues or accommodating the rapid spatial reconfiguration characteristic of surgical environments. Furthermore, the rigorous evaluation of these methods has been constrained by the lack of comprehensive datasets that capture multiple surgical procedures and challenging intraoperative events.

Several specific challenges remain unaddressed in current approaches. First, error accumulation in motion estimation over long sequences becomes particularly problematic during surgical procedures where camera movements and tissue manipulations constantly alter the visual scene. Second, mask misalignment due to camera movement and tissue deformation causes magnification to target incorrect regions when using static masks. Third, boundary artifacts at the interface between magnified and unmagnified regions create visually distracting discontinuities that undermine clinical utility. Fourth, current methods struggle with occlusion handling during instrument intervention, smoke generation, and tissue manipulation, where vessels temporarily disappear and reappear. Finally, existing approaches lack biomechanical modeling of tissue interactions, where vascular pulsations induce variable displacements in surrounding tissues based on their elasticity and connectivity.

To address these challenges, we propose \textbf{EndoControlMag}, a training-free, Lagrangian-based framework with mask-conditioned controllability specifically designed for vascular motion magnification in endoscopic surgery. Our approach introduces two key designs that fundamentally advance the state of the art. First, \textit{Periodic Reference Resetting (PRR)} dynamically updates reference frames at optimal intervals to prevent error accumulation while maintaining temporal coherence across the surgical video sequence. Second, \textit{Hierarchical Tissue-aware Magnification (HTM)} combines dynamic mask tracking with biomechanically-informed adaptive softening to ensure precise, artifact-free magnification that respects the deformable nature of biological tissues. Alongside these technical innovations, we introduce the \textbf{EndoVMM24} dataset, a comprehensive benchmark encompassing multiple surgical specialties and challenging scenarios, enabling a more robust evaluation of magnification techniques under realistic conditions.

Unlike previous approaches that rely heavily on training data or static conditioning, our method leverages off-the-shelf models while introducing novel algorithms optimized for the unique demands of endoscopic surgical scenes. By enabling robust, context-aware vascular motion magnification across diverse surgical procedures, evaluated on a challenging new dataset, EndoControlMag facilitates a surgeon-in-the-loop workflow that enhances both clinical utility and procedural safety.

\begin{figure*}[ht]
    \centering
    \includegraphics[width=1\linewidth]{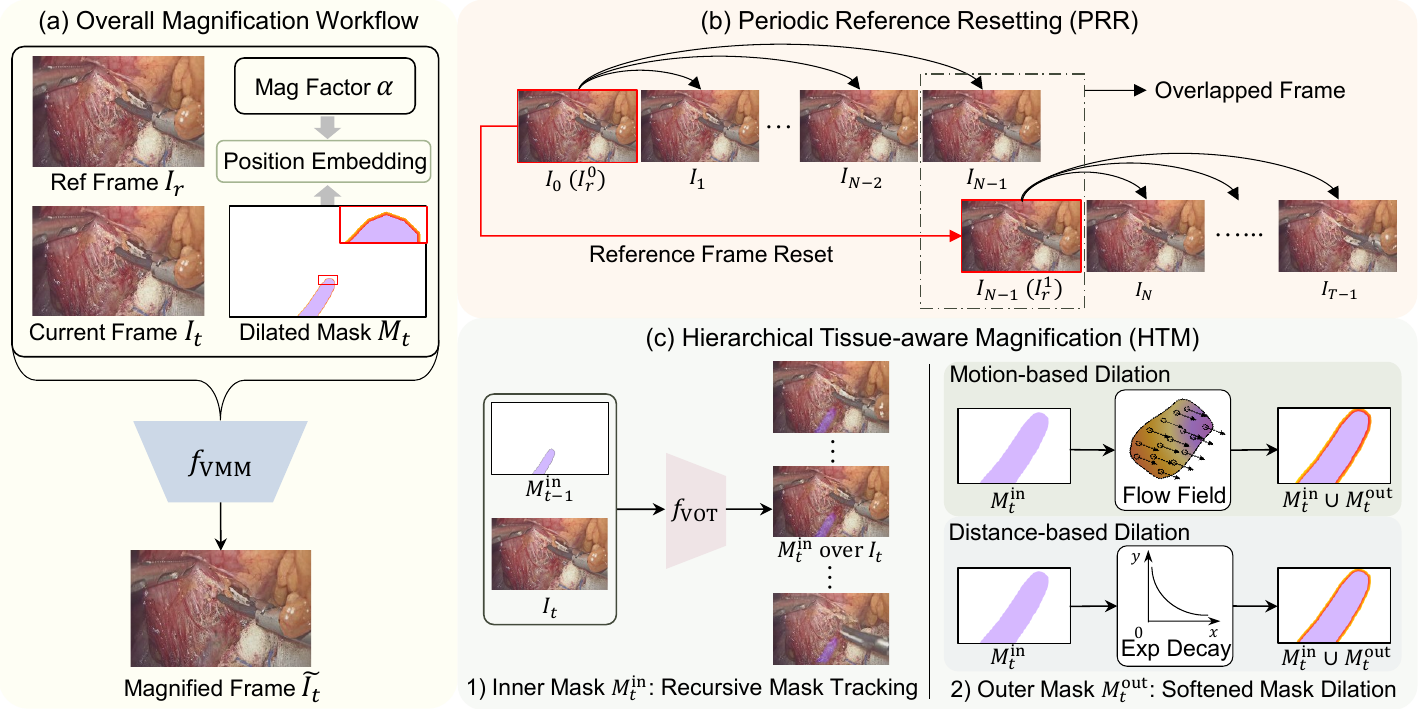}
    \caption{\textbf{Our EndoControlMag for vascular motion magnification in endoscopic surgical video.} 
    (a) The overall magnification pipeline, where vascular motion in the current frame $I_t$ is magnified by the video motion magnification model $f_{\text{VMM}}$ to produce the magnified frame $\tilde{I_t}$. This process utilizes a corresponding reference frame $I_r$ and a hierarchical magnification mask $M_r$, which is positionally encoded by the magnification factor $\alpha$. 
    (b) The Periodic Reference Resetting (PRR) mechanism, which adaptively designates reference frames at regular intervals $N$ with one-frame overlap. This approach prevents cumulative error propagation while maintaining temporal coherence across the surgical video sequence. 
    (c) Our Hierarchical Tissue-aware Magnification (HTM) operates in two stages: 1) recursive updating of the inner mask $M_t^{\text{in}}$ using a pretrained visual object tracking model $f_{\text{VOT}}$ to maintain accurate vessel localization despite view changes and occlusions, and 2) generation of the outer mask $M_t^{\text{out}}$ through adaptive dilation and softening, guided either by optical flow of the surrounding tissue or distance-based exponential decay from the vessel edge. This dual-mask strategy ensures smooth transitions between magnified and unmagnified regions while accommodating the deformable nature of vascular structures and surrounding tissues.}
    \label{fig:overview}
\end{figure*}

\section{Methodology}

\subsection{Preliminaries}\label{sec:pre}
 
\subsubsection{Motion Representation and Lagrangian Magnification}
Accurate motion representation is fundamental for motion magnification. Optical flow, a vector field capturing pixel-level displacements between consecutive frames, forms the basis of Lagrangian approaches. Let $V=\{I_0, I_1, ..., I_{T-1}\}$ denote a video sequence with frames $I_t \in \mathbb{R}^{H \times W \times 3}$, where $H$ and $W$ represent height and width. The optical flow $\mathcal{O}_t \in \mathbb{R}^{H \times W \times 2}$ between frame $I_{t-1}$ and $I_t$ is defined such that:
\begin{equation}
    I_t(\mathbf{x}+\mathcal{O}_t(\mathbf{x})) = I_{t-1}(\mathbf{x}), \quad \forall \mathbf{x} \in \mathbb{R}^{H \times W},
\end{equation}
where $\mathbf{x} = (x, y)$ denotes pixel coordinates in frame $I_{t-1}$, and $\mathcal{O}_t(\mathbf{x}) = (u_t(\mathbf{x}), v_t(\mathbf{x}))$ encodes the horizontal and vertical displacements.

Lagrangian magnification amplifies these displacements over time. For a reference frame $I_r$ and current frame $I_t$, the optical flow $\mathcal{O}_{r \rightarrow t}$ maps pixels from $I_r$ to $I_t$. By scaling $\mathcal{O}_{r \rightarrow t}$ by a factor $\alpha$, the magnified frame $\tilde{I_t}$ is synthesized via backward warping:
\begin{equation}
    \tilde{I_t}(\mathbf{x}) = I_r(\mathbf{x} + \alpha \cdot \mathcal{O}_{r \rightarrow t}(\mathbf{x})).
\end{equation}
This amplifies subtle motions while preserving structural coherence, critical for deformable tissues in endoscopic scenes.

\subsubsection{EndoControlMag Overview}
Our framework, \textbf{EndoControlMag}, extends Lagrangian magnification by integrating hierarchical tissue-aware mask-conditioned control. As illustrated in Fig.~\ref{fig:overview}, let $f_{\text{VMM}}$ denote the base motion magnification model (e.g., FlowMag~\cite{pan2024self}), which generates the magnified frame $\tilde{I_t}$ of current frame $I_t$, based on a magnification mask $M_t$ and the periodically-reset reference frame $I_r$. The mask $M_t$, encoding spatially varying magnification strength with the magnification factor $\alpha$, is derived from vessel recursive tracking and softened dilation, as shown in Fig.~\ref{fig:overview}(c). The process is governed by:
\begin{equation}
    \tilde{I_t} = f_{\text{VMM}}(I_r, I_t, \mathrm{PE}(\alpha) \odot M_t),
\end{equation}
where $\mathrm{PE}(\cdot)$ denotes positional embedding~\cite{mildenhall2021nerf} to encode magnification strength $\alpha$, and $\odot$ represents element-wise multiplication. This formulation enables targeted amplification of vascular motions while suppressing artifacts in static regions.

\subsection{Periodic Reference Resetting}

Accurate motion representation in endoscopic videos is fundamental for Lagrangian-based magnification. While EndoControlMag builds upon optical flow-based motion estimation, we identified that the choice of reference frame significantly influences magnification quality. Prior methods like FlowMag~\cite{pan2024self} anchor magnification to a fixed reference frame $I_0$, which leads to accumulating inaccuracies as the temporal distance $\Delta_t = t - 0$ increases. This becomes particularly problematic in surgical scenarios where camera movements, tissue deformations, and occlusions occur frequently.

To address this limitation, we propose the \textit{Periodic Reference Resetting (PRR)} scheme, which strategically segments the video into overlapping clips with dynamically updated reference frames. This approach effectively bounds cumulative errors in motion estimation while maintaining temporal coherence. Given a video sequence $\mathcal{V} = \{I_0, I_1, \ldots, I_{T-1}\}$, we partition it into $K = \lceil \frac{T}{N-1} \rceil$ clips, where each clip $\mathcal{C}_k = \{I_{s_k}, I_{s_k + 1}, \ldots, I_{s_k + N - 1}\}$ contains $N$ frames with one-frame overlap between consecutive clips. The start index $s_k$ and reference frame $I_r^{k}$ for clip $\mathcal{C}_k$ are defined as:
\begin{equation}
    s_k = k(N-1), \quad I_r^{k} = I_{s_k}, \quad k \in \{0, 1, \ldots, K-1\}.
\end{equation}

Through extensive ablation studies (Sec.~\ref{abl:N}), we empirically determined that $N=4$ provides the optimal balance between error minimization and temporal coherence. With this value, consecutive clips would be structured as $\mathcal{C}_0 = \{I_0, I_1, I_2, I_3\}$, $\mathcal{C}_1 = \{I_3, I_4, I_5, I_6\}$, and so forth. This arrangement ensures that each clip's reference frame $I_r^{k}$ coincides with the last frame of the previous clip $\mathcal{C}_{k-1}$, facilitating smooth transitions across clip boundaries.

The PRR framework incorporates two key design principles.
First, \textit{non-consecutive clip anchoring} creates an intentional overlap between consecutive clips by defining anchor points as $s_k = k(N-1)$ rather than $s_k = kN$. This overlap ensures smooth transitions between reference frames and prevents discontinuities in the magnified motion that would otherwise create jarring visual artifacts.
Second, \textit{error-resettable windows} periodically reset the reference frame to $I_r^{k} = I_{s_k}$, effectively bounding error propagation within $N$-frame windows. This limits the maximum temporal distance between any frame and its reference to $\Delta_t \leq N - 1$, constraining cumulative optical flow errors to $\mathcal{O}(N)$ rather than the $\mathcal{O}(T)$ complexity inherent in fixed-reference approaches.

PRR addresses the unique challenges of endoscopic workflows through its adaptive and continuous design. In surgical settings, rapid camera movements during instrument repositioning induce large \( \Delta_t \)-dependent errors in fixed-reference methods. PRR mitigates this issue by localizing errors within short clips, where optical flow priors remain stable. Additionally, the overlapping clip structure ensures motion continuity, eliminating abrupt transitions between reference anchors and preserving coherence for pulsations that span multiple clips. In conclusion, our approach to reference management enables high-fidelity magnification, effectively addressing the dynamic and unpredictable nature of surgical scenes, where static reference frames are inherently insufficient.

\subsection{Hierarchical Tissue-aware Magnification}

Conventional motion magnification methods often amplify motions uniformly across regions of interest, disregarding the biomechanical relationship between active vascular pulsations and passive tissue displacements. Our framework addresses this limitation through a \textbf{spatially modulated magnification} strategy, guided by a hierarchical dual-mask design that integrates positional encoding for adaptive amplification.

\subsubsection{Inner Mask Recursive Tracking} 
The core vascular structure targeted for magnification is represented by an inner binary mask $M_t^{\text{in}}$, which must be accurately tracked throughout the video sequence to ensure precise spatial control. For the first frame, this mask is initialized as $M_0$ either through manual annotation by an expert surgeon to enable interactivity or via automatic segmentation using vessel-specific models like SurgNet~\cite{chen2023surgnet}. 

As the surgical scene evolves with camera movements, tissue deformation, and instrument interactions, maintaining precise alignment between the magnification region and the target vessel becomes critical for artifact-free visualization. Traditional approaches that rely on static masks inevitably lead to misalignments and visual artifacts when the vessel changes position relative to the camera view.
To address this challenge, we employ a video object tracking (VOT) module $f_{\text{VOT}}$ that propagates the vessel mask recursively through the sequence:
\begin{equation}
    M_t^{\text{in}} = f_{\text{VOT}}(I_t, M_{t-1}^{\text{in}}).
\end{equation}

This recursive formulation ensures temporal consistency by leveraging the mask from the previous frame $M_{t-1}^{\text{in}}$ to predict the current mask $M_{t}^{\text{in}}$ based on the current frame $I_t$. We implement $f_{\text{VOT}}$ using MFT~\cite{neoral2024mft}, a long-term point tracking model specifically designed for challenging scenarios with occlusions and view changes. Unlike conventional object tracking methods~\cite{cheng2022xmem} that track based on semantic features and struggle with the deformable nature of vascular structures, we track individual points within the mask boundary, providing more resilient performance during the complex vessel deformations typical in surgical scenes.

The dynamic mask updating mechanism provides three critical advantages over fixed-mask approaches used in previous works like FlowMag~\cite{pan2024self}: view-invariant magnification that accommodates camera movement, boundary precision that prevents misalignment artifacts, and occlusion resilience that maintains coherence despite temporary instrument obstruction. This tracking-based adaptation is particularly valuable during complex surgical maneuvers where maintaining vessel visibility is critical for intraoperative decision-making.

\subsubsection{Outer Mask Softened Dilation} 
To model the biomechanical interaction between vascular structures and surrounding tissues, we generate an outer region $M_t^{\text{out}}$ through adaptively scaled morphological dilation:
\begin{equation}
    M_t = M_t^{\text{in}} \oplus \mathcal{K}(r), \quad r = \lfloor \gamma \cdot d_{\text{min}}(M_t^{\text{in}}) \rfloor,
\end{equation}
where ``$\oplus$'' stands for morphological dilation operation, $\mathcal{K}(r)$ represents a circular structuring element with radius $r$, ``$\lfloor \cdot \rfloor$'' ensures an integer radius for the dilation kernel, and $d_{\text{min}}(M_t^{\text{in}})$ denotes the minimum distance from the vessel's centroid to its boundary. Unlike FlowMag~\cite{pan2024self}, which uses a fixed dilation radius, our approach adaptively scales the dilation based on the vessel's dimensions with scaling factor $\gamma = 1/15$. This ensures that the transition region remains proportional to vessel size across different anatomical structures. This adaptive scaling is particularly important in surgical scenarios where vessel diameters vary significantly, allowing smaller vessels to have appropriately smaller transition regions while larger vessels receive proportionally wider dilation zones.
We define $M_t^{\text{out}} = M_t \setminus M_t^{\text{in}}$ (where ``$\setminus$'' represents set subtraction) and implement two complementary softening strategies for the transition of magnification strength $W_t$.

For \textbf{motion-based softening}, we explicitly model tissue displacement induced by vascular pulsation using optical flow between the reference frame $I_r^{k}$ (from PRR) and the current frame $I_t$:
\begin{equation}
    \mathcal{O}_t(\mathbf{x}) = \mathcal{F}\left(I_r^{k}, I_t\right)(\mathbf{x}), \quad \forall \mathbf{x} \in M_t^{\text{out}},
\end{equation}
where $\mathcal{F}(\cdot)$ is the flow estimator RAFT~\cite{teed2020raft}. The motion-based weights are derived from normalized flow magnitudes:
\begin{equation}
    W_t^{\text{mot}}(\mathbf{x}) = \frac{\|\mathcal{O}_t(\mathbf{x})\|_2}{\max\limits_{\mathbf{y} \in M_t^{\text{out}}} \|\mathcal{O}_t(\mathbf{y})\|_2}. \label{eq:w_mot}
\end{equation}
This normalization ensures $W_t^{\text{mot}}(\mathbf{x}) \in [0, 1]$, with magnification strength proportional to observed tissue displacement.

For scenarios where optical flow estimation is unreliable (e.g., electrocautery smoke, rapid instrument occlusion), we design the \textbf{distance-based softening}, grounded in the biomechanical attenuation of vascular pulsations in deformable tissues. Let \( d(\mathbf{x}, \partial M_t^{\text{in}}) \) denote the Euclidean distance from pixel \( \mathbf{x} \) to the boundary of the inner vascular mask \( \partial M_t^{\text{in}} \). The distance-based weight \( W_t^{\text{dist}}(\mathbf{x}) \) decays exponentially with this distance:
\begin{equation}
    W_t^{\text{dist}}(\mathbf{x}) = e^{-\beta \cdot d(\mathbf{x}, \partial M_t^{\text{in}})}, \label{eq:w_dist}
\end{equation}
where \( \beta \) controls the attenuation rate, empirically set to 1 to approximate the viscoelastic damping observed in soft tissues. 
The exponential decay ensures stronger magnification near the vascular boundary with gradual reduction in peripheral regions, mimicking the natural viscoelastic damping observed in biological tissues

\begin{algorithm}[t]
\caption{EndoControlMag Workflow.}
\label{alg:endocontrolmag}
\begin{algorithmic}[1]
\REQUIRE Video sequence $\mathcal{V} = \{I_0, \dots, I_{T-1}\}$ \\
\phantom{\quad} Initial vessel mask $M_0$ \\
\phantom{\quad} Clip length $N$, Magnification factor $\alpha$ \\
\phantom{\quad} Dilation ratio $\gamma$, Decay rate $\beta$, Softening mode
\ENSURE Magnified video $\tilde{\mathcal{V}} = \{\tilde{I}_0, \dots, \tilde{I}_{T-1}\}$
\STATE Initialize $k \gets 0$, $\tilde{\mathcal{V}} \gets \emptyset$, $M_0^{\text{in}} \gets M_0$
\STATE Set initial reference frame $I_r^{0} \gets I_0$
\FOR{$t \gets 0$ \TO $T-1$}
    \IF{$t \equiv 0 \pmod{N-1} \land t \neq 0$}
        \STATE $k \gets k + 1$ \COMMENT{Update clip index}
        \STATE $I_r^{k} \gets I_t$ \COMMENT{PRR reference update}
    \ENDIF
    \STATE Track vessel mask: $M_t^{\text{in}} \gets f_{\text{VOT}}(I_t, M_{t-1}^{\text{in}})$
    \STATE Calculate dilation kernel radius: $r \gets \lfloor\gamma\cdot d_{\text{min}}(M_t^{\text{in}})\rfloor$
    \STATE Compute unified mask: $M_t \gets M_t^{\text{in}} \oplus \mathcal{K}(r)$
    \STATE Extract outer mask: $M_t^{\text{out}} \gets M_t \setminus M_t^{\text{in}}$
    \STATE Compute magnification strength $W_t$:
        \IF{softening mode = ``motion''}
            \STATE Estimate optical flow: $\mathcal{O}_t \gets \mathcal{F}(I_r^{k}, I_t)$ 
            \STATE $W_t(\mathbf{x}) \gets \frac{\|\mathcal{O}_t(\mathbf{x})\|_2}{\max_{\mathbf{y}\in M_t^{\text{out}}} \|\mathcal{O}_t(\mathbf{y})\|_2}$ \COMMENT{Eq.~\ref{eq:w_mot}}
        \ELSIF{softening mode = ``distance''}
            \STATE $W_t(\mathbf{x}) \gets \exp(-\beta \cdot d(\mathbf{x},\partial M_t^{\text{in}}))$ \COMMENT{Eq.~\ref{eq:w_dist}}
        \ENDIF
    \STATE Build unified magnification map: 
    \[
    M_t(\mathbf{x}) \gets \begin{cases}
    1, & \mathbf{x} \in M_t^{\text{in}}; \\
    W_t(\mathbf{x}), & \mathbf{x} \in M_t^{\text{out}}; \\
    0, & \text{otherwise}.
    \end{cases}
    \]
    \STATE Synthesize frame: $\tilde{I_t} \gets f_{\text{VMM}}(I_r^{k}, I_t, \text{PE}(\alpha) \odot M_t)$
    \STATE $\tilde{\mathcal{V}}.\text{append}(\tilde{I_t})$
\ENDFOR
\RETURN $\tilde{\mathcal{V}}$
\end{algorithmic}
\end{algorithm}

Herein, the unified magnification map \( M_t \) combines both regions can be represented as:
\begin{equation}
    M_t(\mathbf{x}) = \begin{cases} 
    1, & \mathbf{x} \in M_t^{\text{in}}; \\
    W_t(\mathbf{x}), & \mathbf{x} \in M_t^{\text{out}}; \\
    0, & \text{otherwise}.
    \end{cases}
\end{equation}
Our Hierarchical Tissue-aware Magnification (HTM) framework introduces three critical advancements tailored to surgical dynamics. First, dynamic tracking ensures continuous alignment with deforming vessels despite camera movements. Second, the outer region explicitly models biomechanical wave propagation, distinguishing active vascular motions from passive tissue responses. Third, softening strategies adapt to surgical context: motion-guided weights prioritize regions displaced by pulsation, while distance-based weights mimic natural mechanical attenuation. Our comprehensive approach ensures biomechanically plausible magnification that preserves anatomical relationships while enhancing vascular visibility throughout diverse surgical scenarios.
The complete implementation workflow is formally described in Algorithm~\ref{alg:endocontrolmag}.

\begin{figure*}[t]
\centering
\includegraphics[width=\linewidth, trim=0 12 0 0,clip]{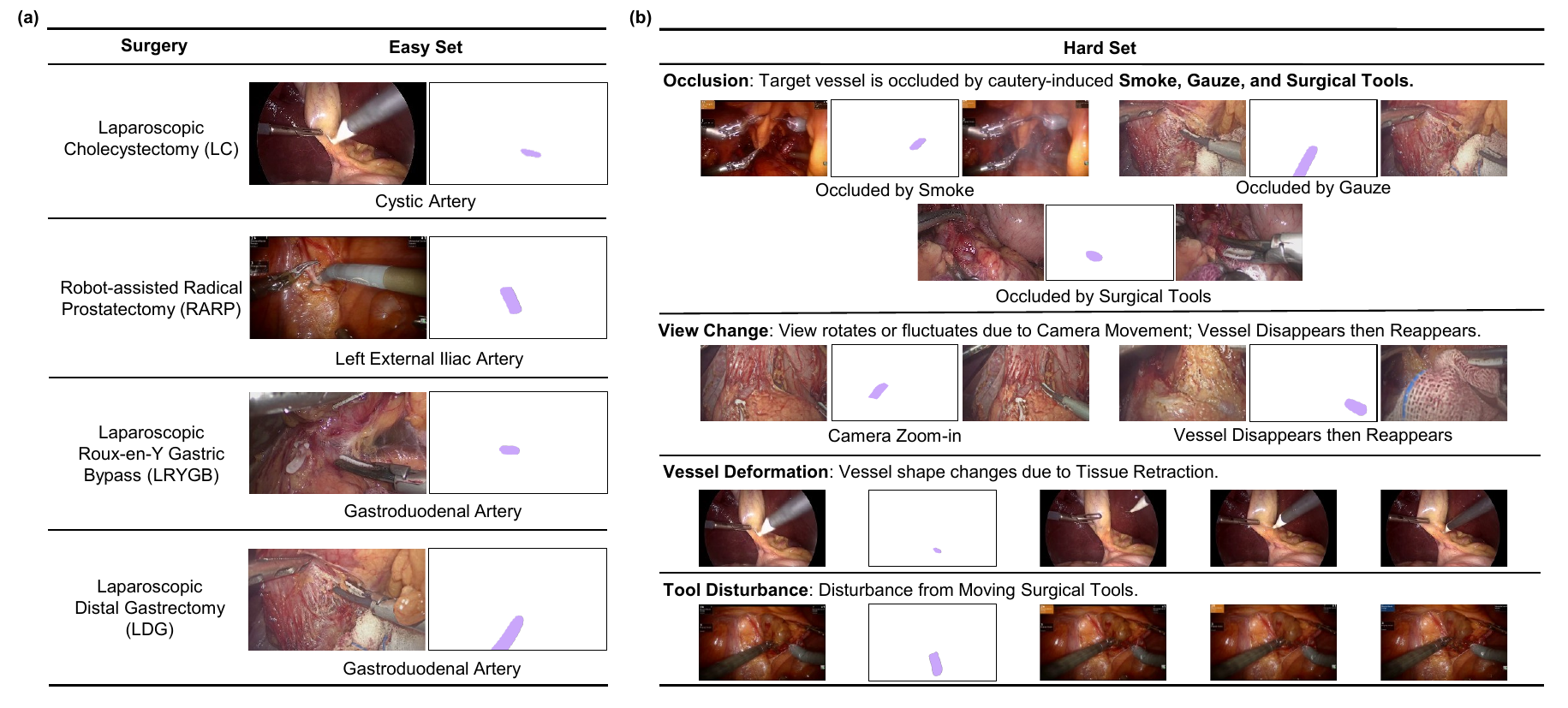}
\caption{\textbf{The composition of our \textbf{EndoVMM24} dataset spanning multiple surgical specialties and challenging clinical scenarios.} (a) The \textit{Easy Set} contains representative clips from four distinct surgical procedures (Cholecystectomy, Prostatectomy, Gastric Bypass, and Laparoscopic Distal Gastrectomy), each featuring a specific vascular structure with minimal movement relative to the camera. The vessel masks (purple overlay) indicate the regions targeted for magnification. (b) The \textit{Hard Set} comprises video clips systematically categorized into four surgical challenges: Occlusion (vessels temporarily obscured by cautery-induced smoke, gauze, or surgical tools), View Change (camera movement, vessel disappearance then reappearance), Vessel Deformation (morphological changes due to tissue retraction), and Tool Disturbance (tool-tissue interaction affecting vessel visualization). This comprehensive dataset enables rigorous evaluation of magnification algorithms across diverse real-world surgical conditions.}
\label{fig:dataset}
\end{figure*}

\begin{table}[t]
  \centering
  \caption{\textbf{Comparison of datasets used for evaluating endoscopic vascular motion magnification.} Our proposed EndoVMM24 dataset significantly surpasses prior works~\cite{janatka2018higher, janatka2020surgical, fan2021robotically, yang2024magnification} in terms of procedural diversity (4 surgery types vs. 1-2), data volume (24 video clips vs. 1-7), and explicit inclusion of challenging surgical scenarios (4 challenge types vs. 0-1), enabling more comprehensive and robust algorithm evaluation.}
  \resizebox{\linewidth}{!}{
    \begin{tabular}{lccc}
    \toprule
    Method & \multicolumn{1}{l}{Surgery Types} & \multicolumn{1}{l}{Total Video Clips} & \multicolumn{1}{l}{Challenge Types} \\
    \midrule
    Janatka et al.~\cite{janatka2018higher} & 1     & 1     & 0 \\
    TMASF~\cite{janatka2020surgical} & 1     & 4     & 1 \\
    Fan et al.~\cite{fan2021robotically}  & 1     & 7     & 0 \\
    AH-PVM~\cite{yang2024magnification} & 2     & 4     & 0 \\
    EndoVMM24 (Ours) & 4     & 24    & 4 \\
    \bottomrule
    \end{tabular}%
    }
  \label{tab:data_comp}%
\end{table}%

\section{Experimental Settings}
\subsection{Dataset}

The rigorous evaluation of vascular motion magnification algorithms has been hampered by limitations in existing datasets, which often fail to capture the full complexity and variability of real-world surgical environments. As summarized in Table~\ref{tab:data_comp}, prior studies have typically relied on datasets with restricted scope, such as those focusing on only one or two surgical procedures~\cite{janatka2018higher, janatka2020surgical, fan2021robotically, yang2024magnification}, incorporating a limited number of video clips (ranging from 1 to 7), and neglecting or only minimally addressing challenging surgical scenarios (0 or 1 challenge type included). Specific limitations include an overemphasis on close-up cropped frames~\cite{yang2024magnification} that disregard global scene dynamics, assumptions of static vessel positioning~\cite{janatka2018higher,fan2021robotically} that ignore camera and tissue movement, or a narrow focus on single procedures~\cite{janatka2020surgical} that restricts generalizability. To overcome these deficiencies and facilitate a more comprehensive assessment of algorithm robustness across diverse clinical conditions, we curated \textbf{EndoVMM24} (Endoscopic Vascular Motion Magnification of 24 video clips). This multi-procedure dataset significantly expands upon previous work by encompassing four distinct surgical specialties and explicitly including four types of challenging scenarios, as illustrated in Fig.~\ref{fig:dataset}, thereby providing a more realistic and demanding benchmark for evaluating endoscopic VMM techniques.

The EndoVMM24 dataset comprises vascular recordings from four distinct surgical domains: Laparoscopic Cholecystectomy (LC) procedures from Cholec80~\cite{rios2023cholec80}, focusing on cystic artery visualization; Robot-assisted Radical Prostatectomy (RARP) procedures from GraSP~\cite{ayobi2024pixel}, highlighting the left external iliac artery; Laparoscopic Roux-en-Y Gastric Bypass (LRYGB) from MultiBypass140~\cite{ramesh2023weakly}, capturing gastroduodenal artery; and Laparoscopic Distal Gastrectomy (LDG) procedures from Nanfang Hospital, featuring the common hepatic artery and gastroduodenal artery. This deliberate anatomical and procedural diversity ensures algorithm evaluation across varying vascular morphologies, tissue characteristics, and surgical workflows—factors critical for clinical generalizability.

We structured the EndoVMM24 dataset into two complementary subsets to facilitate systematic performance analysis. The \textit{Easy Set} includes one representative clip from each procedure (4 in total), showcasing relatively stable vessel positioning with minimal camera movement. This subset provides a controlled environment analogous to conditions addressed in prior work and serves as our baseline for comparative analysis across all methods. The \textit{Hard Set} consists of 20 video clips strategically categorized into four surgical challenges:

\begin{itemize}
    \item \textbf{Occlusion} (8 clips): Vessels temporarily obscured by cautery-induced smoke, surgical gauze, or instrument interventions—scenarios that test algorithm robustness to temporary target disappearance.
    
    \item \textbf{View Change} (6 clips): Camera rotations, zoom operations, and vessel disappearance/reappearance events that challenge spatial continuity in magnification.
    
    \item \textbf{Vessel Deformation} (3 clips): Morphological alterations of vascular structures during tissue retraction and manipulation, requiring algorithms to adapt to changing target shapes.
    
    \item \textbf{Tool Disturbance} (3 clips): Direct tool-tissue interactions adjacent to vessels that create complex motion patterns and potential occlusions.
\end{itemize}

This structured categorization enables quantitative assessment of algorithm performance under increasingly challenging conditions, providing insights into robustness and clinical applicability across the spectrum of real-world surgical scenarios. Unlike previous datasets that focused primarily on ideal conditions, EndoVMM24 deliberately incorporates the visual challenges routinely encountered in clinical practice, allowing for a more realistic evaluation of the potential utility of magnification algorithms in surgical workflows.

\subsection{Implementation Details}
\subsubsection{Baselines}\label{sec:baseline}
We compare our approach primarily with FlowMag~\cite{pan2024self}, the foundation model upon which we build targeted optimizations for surgical settings. Additionally, we include comparisons with a classic traditional motion magnification method, Eulerian Video Magnification (EVM)~\cite{wu2012eulerian}, and several deep learning-based methods, including DMM~\cite{oh2018learning}, the D1 and D2 variants of MDL-VMM~\cite{singh2023multi}, and both the static and dynamic modes of STB-VMM~\cite{lado2023stb} and Axial-VMM~\cite{byung2025learning}.
For fair comparison, we generally adopt the default parameters recommended by the original authors of each baseline. We exclude direct comparisons with methods specifically designed for vascular motion magnification~\cite{janatka2018higher,janatka2020surgical,yang2024magnification,zheng2024localization,fan2021robotically} due to the unavailability of their publicly accessible implementations for reproducibility. 
To ensure comprehensive performance evaluation across different magnification intensities, we systematically test all methods using magnification factors ranging from moderate ($\times$2) to extreme ($\times$32) amplification, specifically $\alpha \in \{2^1, 2^2, 2^3, 2^4, 2^5\}$. This range allows us to assess both subtle enhancements suited for routine visualization and stronger magnifications needed for revealing the faintest vascular pulsations in challenging surgical conditions.

\subsubsection{Quantitative Metrics} \label{sec:metric}
To quantitatively evaluate the performance of our method against the baselines, we employ two complementary categories of assessment:

\paragraph{Image Quality}
We assess the structural fidelity and perceptual quality of magnified videos using three established metrics: Structural Similarity Index (SSIM)~\cite{wang2004image}, which measures the preservation of structural information between original and magnified frames; Peak Signal-to-Noise Ratio (PSNR)~\cite{hore2010image}, which quantifies noise levels introduced during magnification; and Multi-Scale Image Quality Transformer (MUSIQ)~\cite{ke2021musiq}, a learning-based perceptual quality metric that evaluates images across multiple scales. These metrics provide complementary perspectives on visual quality, with higher values indicating better results for all three measures.

\begin{table*}[ht]
\caption{
\textbf{Quantitative evaluation of image quality on the Easy Set.} We compare our EndoControlMag against both unconditional and conditional baseline methods across five magnification factors using three metrics: SSIM, PSNR, and MUSIQ. Mean and standard deviation are reported for each metric. Our method consistently outperforms baselines on structural preservation metrics, with both variants achieving superior SSIM and PSNR scores across all magnification factors, while maintaining competitive perceptual quality. Best results are highlighted in \textbf{bold}, with runner-up results \underline{underlined}.
}
\resizebox{\textwidth}{!}{ 
\centering
\begin{tabular}{llllllllllllllll}
\toprule
\multirow{2}[1]{*}{Method}           & \multicolumn{5}{c}{SSIM $\uparrow$}                                                                                    & \multicolumn{5}{c}{PNSR $\uparrow$}                                                                                         & \multicolumn{5}{c}{MUSIQ $\uparrow$}                                                                                           \\ \cmidrule(lr){2-6} \cmidrule(lr){7-11} \cmidrule(lr){12-16}
         & $\times$2                  & $\times$4                  & $\times$8                  & $\times$16                 & $\times$32                 & $\times$2                   & $\times$4                   & $\times$8                   & $\times$16                  & $\times$32                  & $\times$2                    & $\times$4                    & $\times$8                   & $\times$16                   & $\times$32                  \\ \midrule
\multicolumn{16}{l}{Unconditional Magnification}   
\\ \midrule
EVM~\cite{wu2012eulerian}              & 0.81$\pm$0.03          & 0.76$\pm$0.05          & 0.76$\pm$0.10          & 0.79$\pm$0.09          & 0.68$\pm$0.10          & 25.87$\pm$0.58          & 24.18$\pm$1.78          & 22.74$\pm$1.52          & 20.80$\pm$2.09          & 20.10$\pm$2.84          & 41.79$\pm$7.77           & 38.04$\pm$8.44           & 37.91$\pm$8.08          & 32.64$\pm$5.46           & 25.66$\pm$6.80          \\
DMM~\cite{oh2018learning}        & 0.89$\pm$0.04          & 0.80$\pm$0.07          & 0.76$\pm$0.08          & 0.74$\pm$0.09          & 0.72$\pm$0.09          & 30.26$\pm$1.53          & 26.18$\pm$2.23          & 23.93$\pm$2.17          & 22.55$\pm$2.15          & 21.82$\pm$2.33          & \textbf{50.41$\pm$8.19}           & \textbf{49.73$\pm$7.64}           & 46.23$\pm$7.79          & 38.55$\pm$9.26           & 30.74$\pm$8.77          \\
MDL-VMM-D1~\cite{singh2023multi}  & 0.83$\pm$0.05          & 0.79$\pm$0.07          & 0.77$\pm$0.07          & 0.76$\pm$0.07          & 0.77$\pm$0.06          & 26.89$\pm$1.61          & 25.02$\pm$1.64          & 23.96$\pm$1.66          & 23.24$\pm$1.75          & 22.99$\pm$1.65          & 38.69$\pm$11.58          & 34.68$\pm$10.06          & 32.02$\pm$8.18          & 28.76$\pm$6.91           & 25.31$\pm$7.28          \\
MDL-VMM-D2~\cite{singh2023multi}  & 0.86$\pm$0.03          & 0.82$\pm$0.05          & 0.79$\pm$0.07          & 0.77$\pm$0.07          & 0.77$\pm$0.07          & 27.80$\pm$0.84          & 25.71$\pm$1.55          & 24.21$\pm$1.82          & 23.19$\pm$1.90          & 22.81$\pm$1.65          & 40.18$\pm$11.57          & 33.44$\pm$8.54           & 28.19$\pm$7.00          & 24.64$\pm$6.72           & 22.84$\pm$6.99          \\
STB-VMM-static~\cite{lado2023stb}   & 0.77$\pm$0.06          & 0.77$\pm$0.08          & 0.79$\pm$0.08          & 0.78$\pm$0.07          & 0.78$\pm$0.06          & 24.52$\pm$1.58          & 23.56$\pm$2.76          & 24.24$\pm$3.15          & 23.77$\pm$2.53          & 23.66$\pm$2.05          & 43.24$\pm$6.65           & 39.65$\pm$4.78           & 36.45$\pm$4.86          & 29.27$\pm$6.97           & 25.56$\pm$8.74          \\
STB-VMM-dynamic~\cite{lado2023stb}  & 0.84$\pm$0.11          & 0.80$\pm$0.07          & 0.75$\pm$0.08          & 0.74$\pm$0.08          & 0.74$\pm$0.09          & 27.39$\pm$4.09          & 25.55$\pm$2.00          & 23.44$\pm$2.19          & 22.18$\pm$2.29          & 21.73$\pm$2.66          & {46.91$\pm$5.77}     & 45.09$\pm$7.05           & {44.28$\pm$6.01} & 44.14$\pm$5.58           & 35.15$\pm$7.64          \\
Axial-VMM-static~\cite{byung2025learning}  & 0.87$\pm$0.03          & 0.81$\pm$0.04          & 0.79$\pm$0.05          & 0.78$\pm$0.05          & 0.77$\pm$0.05          & 29.12$\pm$0.65          & 26.11$\pm$0.45          & 24.55$\pm$0.88          & 23.73$\pm$1.15          & 23.41$\pm$1.11          & 41.11$\pm$7.53           & 36.16$\pm$8.31           & 31.62$\pm$7.60          & 29.17$\pm$6.64           & 26.91$\pm$6.89          \\
Axial-VMM-dynamic~\cite{byung2025learning} & 0.87$\pm$0.04          & 0.80$\pm$0.07          & 0.75$\pm$0.08          & 0.73$\pm$0.09          & 0.70$\pm$0.09          & 28.90$\pm$0.66          & 25.97$\pm$1.48          & 23.91$\pm$1.86          & 22.36$\pm$2.07          & 21.41$\pm$2.31          & 43.90$\pm$7.09           & 43.17$\pm$6.72           & 40.87$\pm$6.22          & 36.02$\pm$7.19           & 29.05$\pm$8.06          \\ \midrule
\multicolumn{16}{l}{Conditional Magnification}
\\ \midrule
FlowMag~\cite{pan2024self}          & \uline{0.96$\pm$0.01}          & \uline{0.96$\pm$0.01}          & {0.95$\pm$0.02}          & {0.95$\pm$0.02}          & {0.95$\pm$0.02}          & {35.51$\pm$1.49}          & {34.65$\pm$1.98}          & {34.13$\pm$2.28}          & {33.70$\pm$2.54}          & {32.57$\pm$2.30}          & 48.24$\pm$11.08          & 46.98$\pm$10.45          & {46.36$\pm$10.16}         & {45.83$\pm$9.81}           & {45.74$\pm$9.59} \\
EndoControlMag (Distance)       & \textbf{0.97$\pm$0.01} & \textbf{0.97$\pm$0.01} & \textbf{0.96$\pm$0.01} & \textbf{0.96$\pm$0.01} & \textbf{0.96$\pm$0.01} & \textbf{36.94$\pm$1.42} & \textbf{36.21$\pm$0.82} & \textbf{35.62$\pm$0.63} & \textbf{35.20$\pm$0.71} & \uline{33.69$\pm$1.05} & {48.95$\pm$11.64} & {48.51$\pm$11.64} & \textbf{47.93$\pm$11.62}   & \uline{47.39$\pm$11.68} & \uline{47.31$\pm$11.49}   \\ 
EndoControlMag (Motion)       & \textbf{0.97$\pm$0.01} & \textbf{0.97$\pm$0.01} & \uline{0.96$\pm$0.02} & \uline{0.96$\pm$0.01} & \uline{0.96$\pm$0.02} & \uline{36.03$\pm$1.76} & \uline{35.71$\pm$1.17} & \uline{35.09$\pm$0.87} & \uline{34.73$\pm$0.92} & \textbf{34.30$\pm$1.05} & \uline{49.37$\pm$11.44} & \uline{48.82$\pm$11.80} & \uline{47.74$\pm$11.75}   & \textbf{48.47$\pm$11.40} & \textbf{47.35$\pm$11.42}   \\\bottomrule
\end{tabular}
}
\label{tab:easy_quality}
\end{table*}

\begin{table*}[ht]
\caption{
\textbf{Quantitative evaluation of magnification accuracy on the Easy Set.} We compare our EndoControlMag against both unconditional and learning-based baseline methods across five magnification factors using two complementary metrics: Motion Error ($E_{motion}$) and Magnification Error ($E_{mag}$), both of which should be minimized. Our method demonstrates superior performance in motion fidelity and magnification accuracy, particularly at lower magnification factors where precision is most critical. Best results are highlighted in \textbf{bold}, with runner-up results \underline{underlined}.
}
\resizebox{\textwidth}{!}{ 
\begin{tabular}{lllllllllll}
\toprule
\multirow{2}[1]{*}{Method}           & \multicolumn{5}{c}{$E_{motion}$ $\downarrow$}                                                                                    & \multicolumn{5}{c}{$E_{mag}$ $\downarrow$}                                                                             \\ \cmidrule(lr){2-6} \cmidrule(lr){7-11} 
           & $\times$2                   & $\times$4                   & $\times$8                   & $\times$16                   & $\times$32                    & $\times$2                   & $\times$4                   & $\times$8                   & $\times$16                   & $\times$32                   \\ 
           \midrule
\multicolumn{11}{l}{Unconditional Magnification}\\ 
\midrule
EVM~\cite{wu2012eulerian}              & 2.025$\pm$0.576          & 3.735$\pm$1.743          & 8.072$\pm$4.076          & 17.139$\pm$8.739          & 35.300$\pm$18.047          & 2.046$\pm$0.099          & 4.952$\pm$0.057          & 8.888$\pm$1.164          & 17.828$\pm$3.248          & {33.740$\pm$3.312}          \\
DMM~\cite{oh2018learning}        & 1.525$\pm$0.304          & 4.094$\pm$0.760          & 9.775$\pm$1.988          & 18.098$\pm$5.938          & 33.634$\pm$18.105          & 2.556$\pm$0.468          & 5.882$\pm$0.668          & 17.875$\pm$1.283         & 34.833$\pm$2.429          & 35.595$\pm$3.849          \\
MDL-VMM-D1~\cite{singh2023multi}  & 1.374$\pm$0.104          & 4.068$\pm$1.335          & 8.089$\pm$3.349          & 16.642$\pm$5.780          & 38.255$\pm$12.456          & 1.703$\pm$1.765          & 4.969$\pm$1.653          & {8.237$\pm$6.455}          & 21.053$\pm$40.824         & 46.875$\pm$13.622         \\
MDL-VMM-D2~\cite{singh2023multi}  & 1.191$\pm$1.153          & 4.701$\pm$2.914          & 8.289$\pm$4.002          & 16.898$\pm$9.119          & 37.050$\pm$16.083          & 2.398$\pm$1.921          & 4.448$\pm$3.720          & 8.945$\pm$7.940          & 19.255$\pm$40.207         & 46.270$\pm$12.823         \\
STB-VMM-static~\cite{lado2023stb}   & 1.523$\pm$1.400          & 4.327$\pm$1.332          & 7.598$\pm$5.219          & 16.482$\pm$7.728          & 35.428$\pm$16.714          & 2.330$\pm$2.075          & 4.931$\pm$2.197          & 10.446$\pm$6.186         & 20.125$\pm$11.495         & 44.395$\pm$13.673         \\
STB-VMM-dynamic~\cite{lado2023stb}  & 1.704$\pm$1.259         & 4.035$\pm$2.067         & 9.131$\pm$5.604         & \textbf{12.940$\pm$11.251}         & \textbf{25.392$\pm$14.414}          & {1.244$\pm$2.254}          & 5.826$\pm$2.935          & 12.300$\pm$6.383         & 15.286$\pm$18.904         & 34.738$\pm$12.215         \\
Axial-VMM-static~\cite{byung2025learning}  & 1.251$\pm$0.409          & 3.750$\pm$1.180          & 9.073$\pm$3.275          & 17.660$\pm$8.043          & 32.938$\pm$17.677          & 3.584$\pm$5.018          & 4.580$\pm$2.968          & 8.840$\pm$8.425          & 17.235$\pm$6.102          & 37.483$\pm$8.802          \\
Axial-VMM-dynamic~\cite{byung2025learning} & 1.633$\pm$0.298          & 4.248$\pm$0.896          & 8.695$\pm$1.986          & 17.845$\pm$4.787          & 33.394$\pm$13.400          & 1.303$\pm$1.553          & {3.196$\pm$1.139}          & 9.785$\pm$1.140          & \textbf{14.074$\pm$1.862}          & 38.780$\pm$2.489          \\ 
\midrule
\multicolumn{11}{l}{Conditional Magnification} \\ 
\midrule
FlowMag~\cite{pan2024self}          & {1.151$\pm$0.540}          & {3.347$\pm$1.566}          & {7.314$\pm$3.709}          & 15.261$\pm$8.457          & 31.255$\pm$17.910          & 1.732$\pm$1.060          & 5.067$\pm$3.095          & 9.965$\pm$5.785          & 17.053$\pm$4.212          & 35.110$\pm$1.055          \\
EndoControlMag (Distance)       & \textbf{0.999$\pm$0.564} & \textbf{3.003$\pm$1.721} & \textbf{7.012$\pm$4.022} & {15.066$\pm$8.667} & {31.136$\pm$17.941} & \textbf{1.033$\pm$0.120} & \textbf{2.978$\pm$0.045} & \uline{6.983$\pm$0.101} & {14.980$\pm$0.207} & \uline{30.970$\pm$0.320} \\ 
EndoControlMag (Motion)       & \uline{1.027$\pm$0.554} & \uline{3.132$\pm$1.876} & \uline{7.231$\pm$4.219} & \uline{15.027$\pm$8.502} & \uline{31.021$\pm$17.755} & \uline{1.054$\pm$0.131} & \uline{3.103$\pm$0.067} & \textbf{6.870$\pm$0.104} & \uline{14.755$\pm$0.190} & \textbf{30.324$\pm$0.307} \\ 
\bottomrule
\end{tabular}
}
\label{tab:easy_error}
\end{table*}

\paragraph{Magnification Accuracy}
To specifically evaluate the precision of motion magnification in preserving the target amplification relationship, we adopt two metrics introduced in FlowMag~\cite{pan2024self}:

\begin{itemize}
    \item \textbf{Motion Error ($E_{motion}$)}: Measures the absolute difference between the target magnified motion and the actual achieved motion:
    \begin{equation}
        E_{motion} = \|\mathcal{O}(I_r, I_t) - \alpha \cdot \mathcal{O}(I_r, \tilde{I_t})\|_1,
    \end{equation}
    where $\mathcal{O}(I_r, I_t)$ and $\mathcal{O}(I_r, \tilde{I_t})$ represent the optical flow from reference frame $I_r$ to the original and magnified frames, respectively. Lower values indicate better adherence to the target magnification factor $\alpha$.
    
    \item \textbf{Magnification Error ($E_{mag}$)}: Evaluates the ratio of magnified to original motion magnitude relative to the target magnification factor:
    \begin{equation}
        E_{mag} = \|\frac{\|\mathcal{O}(I_r, \tilde{I_t})\|_2}{\|\mathcal{O}(I_r, I_t)\|_2 + \epsilon} - \alpha\|_1,
    \end{equation}
    where $\epsilon$ is a small constant (1e-7) added to avoid division by zero. This metric specifically penalizes inconsistent magnification across the frame, which is crucial for maintaining proportional enhancement of vascular pulsations.
\end{itemize}

For both accuracy metrics, we use RAFT~\cite{teed2020raft}, a state-of-the-art optical flow estimator, to compute the flow fields. These metrics provide a rigorous quantitative assessment of how faithfully the magnification algorithms preserve the target amplification relationship across dynamic surgical scenes.

For a more comprehensive understanding of our framework and to appreciate the effectiveness of motion magnification beyond static images, we encourage readers to visit our project website\footnote{\url{https://szupc.github.io/EndoControlMag/}}, which provides video demonstrations and the reference implementation.

\section{Results and Analysis}
To rigorously evaluate the robustness and generalizability of EndoControlMag, we conducted experiments across two subsets of the datasets: the \textit{Easy Set}, designed to assess performance under ideal conditions with minimal surgical complexities, and the \textit{Hard Set}, which introduces real-world challenges such as instrument occlusion, tissue deformation, dynamic view shifts and disturbance from surgical tools. This section presents quantitative and qualitative comparisons on these two subsets, followed by expert surgical assessments and ablation studies that analyze the impact of key design choices in our framework.

\subsection{Evaluation on Easy Set}

The \textit{Easy Set} comprises four videos (one from each surgical procedure) where vessels remain static relative to the endoscopic camera, enabling fair quantification of motion magnification fidelity against all nine baseline methods mentioned in Sec.~\ref{sec:baseline}. 
For each metric outlined in Sec.~\ref{sec:metric}, we report the mean and standard deviation across these four videos to provide a robust statistical assessment.

\subsubsection{Quantitative Evaluation}
We evaluate both structural integrity and motion accuracy of magnified videos across five magnification factors ($\times$2 to $\times$32). Tables~\ref{tab:easy_quality} and \ref{tab:easy_error} present comprehensive results comparing our approach with baseline methods.

\begin{figure*}[!ht]
\centering
\includegraphics[width=\linewidth]{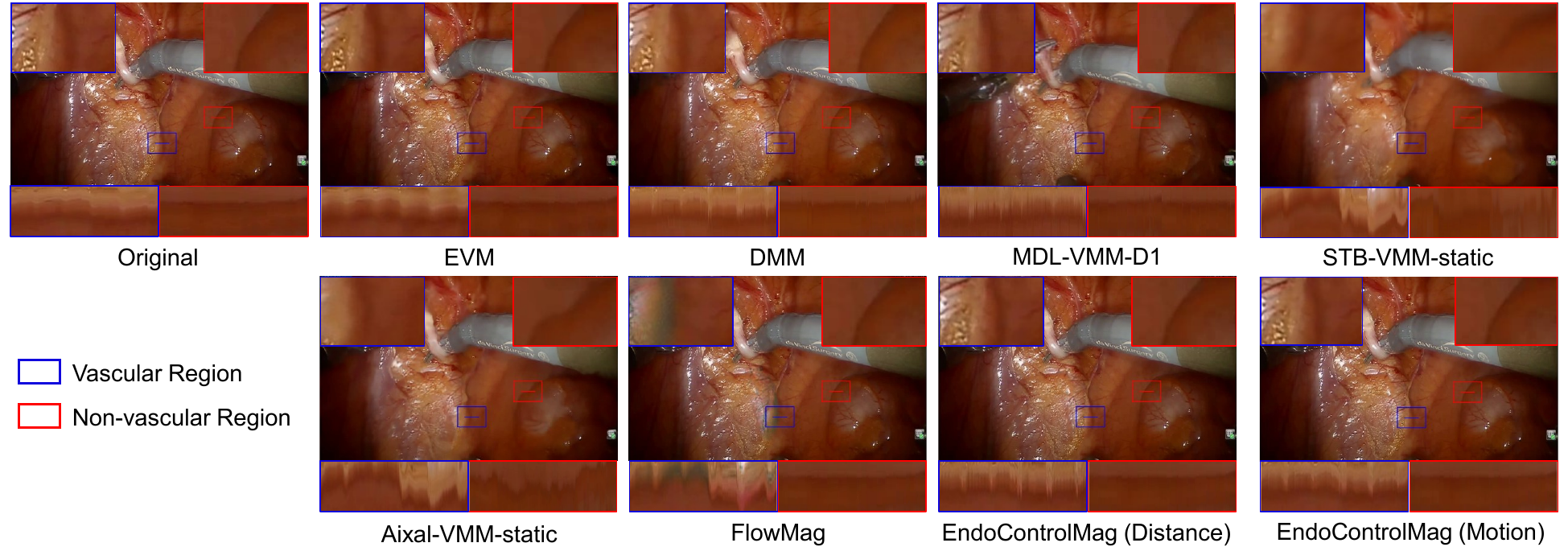}
\caption{\textbf{Qualitative comparison of our EndoControlMag with baseline methods on a representative surgical video with magnification factor $\times 8$.} For each method, we highlight the vascular region (blue box) and non-vascular region (red box) with close-up views shown at the top of each panel. The spatial-temporal slices along the blue and red lines are displayed at the bottom, revealing temporal motion behavior. Compared with other methods, both variants of EndoControlMag achieve superior results with clear vascular motion magnification while maintaining the integrity of non-vascular tissues, demonstrating biomechanically plausible motion amplification with minimal artifacts. The spatial-temporal slices confirm that our method preserves the periodic nature of vascular pulsations while maintaining a stable representation of static tissue regions.}
\label{fig:quality}
\end{figure*}

In terms of image quality, our EndoControlMag consistently outperforms all baseline methods, achieving the highest SSIM and PSNR scores across all magnification strengths. This indicates superior preservation of structural details and effective noise suppression, particularly at higher magnification factors where visual fidelity becomes increasingly critical. The framework's strong MUSIQ scores further confirm its alignment with human perceptual quality. Traditional method EVM~\cite{wu2012eulerian} exhibits significant quality degradation at higher magnifications (PSNR=20.10±2.84 at $\alpha$=32), while unconditioned learning-based approaches introduce visible artifacts due to their reliance on global amplification strategies. Notably, our EndoControlMag variants consistently outperform FlowMag~\cite{pan2024self}, despite both supporting mask-conditioned magnification.

When assessing magnification accuracy through Motion Error ($E_{\text{motion}}$) and Magnification Error ($E_{\text{mag}}$), EndoControlMag demonstrates superior precision in preserving the target amplification relationship. For smaller magnification factors ($\times$2, $\times$4), our method achieves remarkable improvements, with error reductions of 13.2\% in $E_{\text{motion}}$ and 41.2\% in $E_{\text{mag}}$ compared to FlowMag~\cite{pan2024self}. At higher magnification factors, EndoControlMag maintains consistent linear error scaling that follows the theoretical $O(\alpha)$ relationship, while several baseline methods (particularly STB-VMM-dynamic~\cite{lado2023stb} and Axial-VMM-dynamic~\cite{byung2025learning}) exhibit unstable error patterns across different magnification levels.

Both the image quality and magnification accuracy improvements stem from our two key design elements: (1) the PRR scheme, which resets reference frames at optimal intervals to prevent cumulative drift in optical flow estimation, and (2) the HTM framework, which ensures precise vessel tracking while providing smooth transitions at region boundaries. These modules enable EndoControlMag to maintain high-fidelity magnification even at extreme amplification factors, a critical requirement for visualizing subtle vascular pulsations in clinical settings.

\subsubsection{Qualitative Comparison}

Fig.~\ref{fig:quality} presents a visual comparison using a representative surgical video at magnification factor $\times$8. Our evaluation focuses on the effective magnification of the vascular region (blue box) and preservation of the surrounding tissue (red box). The close-up views demonstrate that EndoControlMag successfully magnifies vascular pulsations while suppressing noise in both regions. In contrast, alternative methods introduce visible artifacts such as blur~\cite{oh2018learning,lado2023stb,byung2025learning} and noise~\cite{wu2012eulerian,singh2023multi,pan2024self}.

The spatial-temporal (x-t) slices reveal that our method produces smooth, periodic amplifications that align with cardiac rhythms while preserving the static nature of non-target tissues. Baseline methods exhibit inconsistent fluctuations~\cite{elgharib2015video,singh2023multi,pan2024self} and fail to maintain the stationary properties of non-target tissue borders~\cite{lado2023stb,byung2025learning,oh2018learning}. These visual results confirm the superior capacity of our method for precise, artifact-free motion magnification in surgical videos.

\begin{figure*}[t]
\centering
\includegraphics[width=\linewidth]{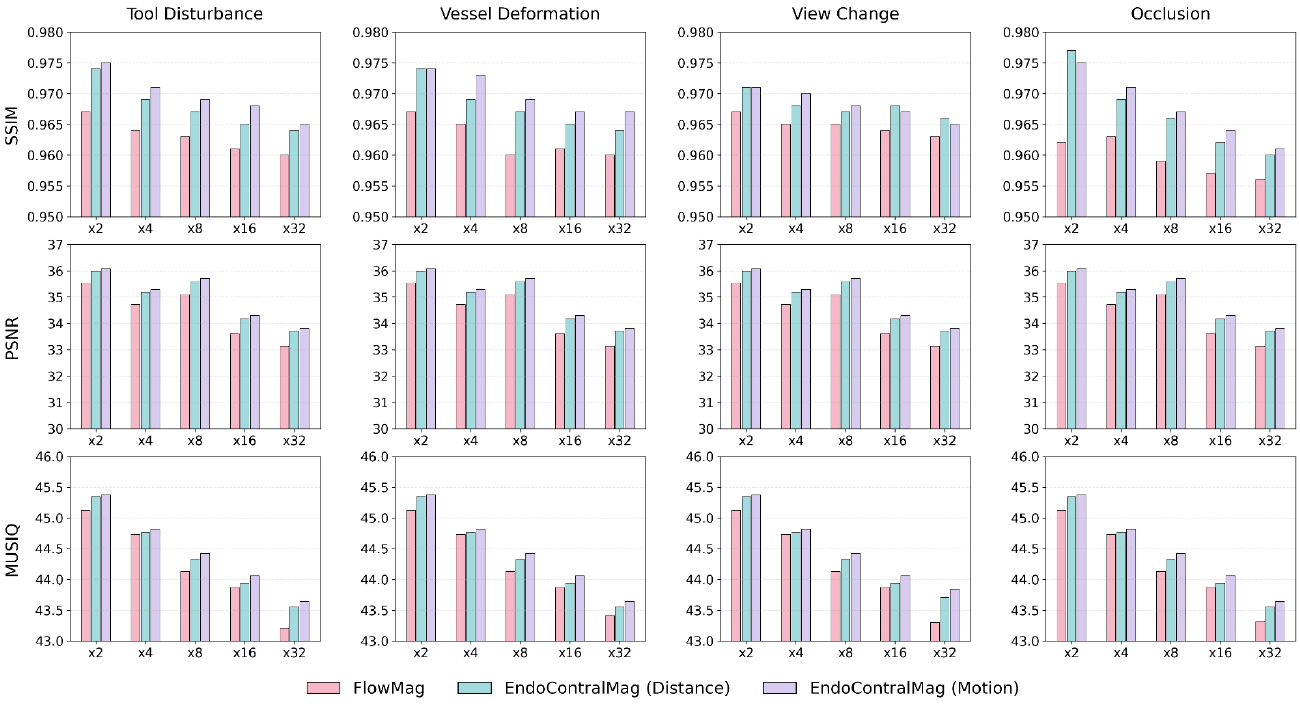}
\caption{\textbf{Quantitative performance comparison of our EndoControlMag against FlowMag~\cite{pan2024self} on the Hard Set across four surgical challenge categories.} Each column represents a specific challenge: Tool Disturbance, Vessel Deformation, View Change, and Occlusion. Results are organized in three rows showing SSIM (top), PSNR (middle), and MUSIQ (bottom) metrics across five magnification factors ($\times$2 to $\times$32). Both EndoControlMag variants consistently outperform FlowMag~\cite{pan2024self} across all metrics and challenge types, with the Motion-based variant showing particular strength in Tool Disturbance and Vessel Deformation scenarios, while both variants maintain robust performance during View Changes and Occlusions.} 
\label{fig:hard_res}
\end{figure*}

\subsection{Evaluation on Hard Set}

The \textit{Hard Set} evaluation represents a critical assessment of magnification robustness under realistic surgical conditions that challenge conventional approaches. Unlike the Easy Set evaluation that compared multiple baseline methods, our Hard Set experiments focus specifically on comparing EndoControlMag with FlowMag~\cite{pan2024self}. This choice stems from their shared capability for mask-conditioned magnification, which is essential for handling challenging surgical scenarios where selective amplification is required while preventing artifact propagation in non-target regions. Unconditional methods applying global magnification would introduce intolerable visual distortions when confronted with the complex dynamics of real surgical environments, rendering them unsuitable for this rigorous evaluation.

Figure~\ref{fig:hard_res} presents a detailed quantitative comparison across four distinct surgical challenge categories: Tool Disturbance, Vessel Deformation, View Change, and Occlusion. Our EndoControlMag framework consistently outperforms FlowMag across all evaluated metrics (SSIM, PSNR, MUSIQ) and within each challenge type, demonstrating its superior robustness and adaptability. Notably, the Motion-based variant generally achieves the highest performance, exhibiting SSIM improvements ranging from 0.5\% to 1.2\% and PSNR gains between 0.53dB and 0.67dB compared to FlowMag. This advantage is particularly pronounced in the ``Tool Disturbance'' and ``Vessel Deformation'' categories, where accurately modeling the complex biomechanical interactions between instruments, tissues, and vessels is paramount.

Delving into specific challenges reveals the mechanisms behind the enhanced performance of our EndoControlMag. During occlusion events, such as temporary vessel obscuration by cautery smoke or instruments, the PRR scheme proves crucial. By dynamically resetting reference frames, it prevents the accumulation of optical flow errors that would otherwise corrupt magnification upon vessel reappearance. This results in more coherent and artifact-free magnification throughout occlusion sequences, as evidenced by the consistently higher SSIM and PSNR values in the ``Occlusion'' column of Fig.~\ref{fig:hard_res}.

For view changes involving camera movement or vessel disappearance and reappearance, the HTM framework's recursive tracking component ($f_{\text{VOT}}$) ensures that the inner magnification mask remains accurately aligned with the target vessel. Simultaneously, the adaptive softening of the outer mask accommodates the changing spatial configuration, preventing abrupt boundary artifacts. This leads to stable performance improvements, as shown in the ``View Change'' column.

In scenarios involving vessel deformation due to tissue retraction or tool manipulation, the Motion-based softening strategy particularly excels. By explicitly modeling tissue displacement patterns derived from optical flow, it adaptively adjusts magnification strength in the transition zone, respecting the biomechanical relationship between the deforming vessel and surrounding tissue. This leads to superior structural preservation, reflected in the higher SSIM scores in the ``Vessel Deformation'' and ``Tool Disturbance'' columns, especially at moderate magnification factors where subtle deformations are prominent.

\begin{figure*}[t]
\centering
\includegraphics[width=\linewidth]{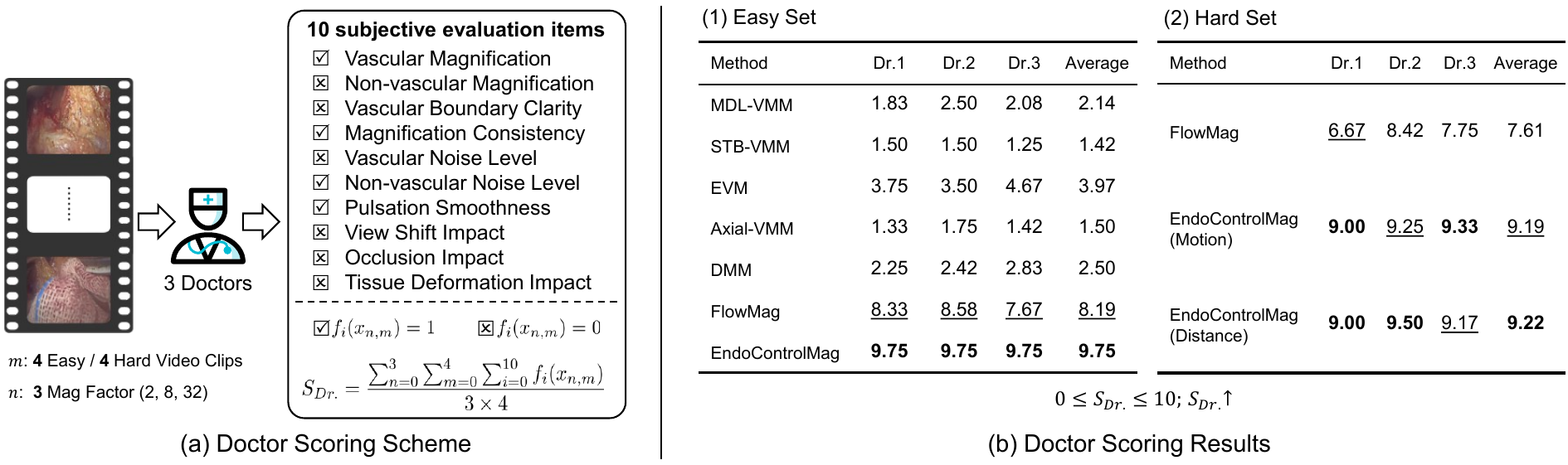}
\caption{\textbf{Clinical evaluation by expert surgeons.} (a) Evaluation methodology: Three experienced surgeons assessed magnification quality using a standardized 10-item scoring rubric covering vascular and non-vascular regions, boundary clarity, consistency, noise levels, pulsation smoothness, and robustness to surgical challenges. Each criterion was evaluated using a binary scoring system (Yes=1, No=0) across 8 videos (4 Easy, 4 Hard) at three magnification factors ($\times$2, $\times$8, $\times$32). (b) Evaluation results: For Easy Set cases, EndoControlMag achieved near-perfect scores (9.75) from all evaluators, significantly outperforming FlowMag~\cite{pan2024self} (8.19) and other baselines. For Hard Set cases, both variants of EndoControlMag (Motion: 9.19, Distance: 9.22) substantially outperformed FlowMag~\cite{pan2024self} (7.61), with strong agreement among all evaluators, demonstrating the clinical superiority of our approach under realistic surgical conditions.}
\label{fig:doctor}
\end{figure*}

The superior performance of EndoControlMag in these demanding scenarios stems from the synergistic interplay between its core components. The PRR scheme effectively bounds temporal error propagation within short, manageable clips, ensuring resilience against large camera movements or transient occlusions that would destabilize fixed-reference methods. Concurrently, the HTM framework, with its precise vessel tracking and adaptive softening strategies, prevents spatial misalignment and ensures biomechanically plausible transitions between magnified and non-magnified regions. This combination confers robustness critical for real-world surgical settings, where conditions change rapidly and unpredictably.

Interestingly, the comparison between our two softening strategies highlights their complementary strengths. Motion-based softening demonstrates a slight edge in scenarios dominated by complex tissue deformation and tool disturbance, leveraging accurate optical flow to model displacement patterns. Conversely, distance-based softening provides more stable and reliable results during occlusions or rapid movements where optical flow estimation can become unreliable, relying instead on a consistent biomechanical decay model. This validates our dual-strategy design, allowing the framework to adapt implicitly or explicitly to varying surgical conditions and ensuring robust performance across a wider range of intraoperative events.

\subsection{Surgeon Evaluation}
To assess the clinical relevance and perceptual quality of the magnification results, we conduct a double-blind evaluation involving three experienced surgeons. As outlined in Fig.~\ref{fig:doctor} (a), the evaluation protocol utilizes a standardized 10-item scoring rubric designed to capture critical aspects of magnification performance in a surgical context. These criteria include the clarity and consistency of vascular magnification, noise levels in both vascular and non-vascular regions, smoothness of pulsation visualization, and robustness to common surgical challenges such as view shifts, occlusions, and tissue deformation.
For the evaluation, we select a representative subset of 8 video clips (4 from the Easy Set, 4 from the Hard Set) from our EndoVMM24 dataset. Each clip is processed by the competing methods at three distinct magnification factors ($\times$2, $\times$8, $\times$32) to assess performance across different amplification levels. To mitigate bias, the study employs a double-blind design where both the identity of the magnification method and the specific video clip are anonymized for the evaluating surgeons. Each surgeon independently assesses the magnified videos using the 10-item rubric, assigning a binary score (Yes=1, No=0) for each criterion. The final score for each method on a given set (Easy or Hard), denoted as $S_{Dr.}$, represents the average score across all criteria, videos within the set, and magnification factors, resulting in a maximum possible score of 10.

The evaluation results, summarized in Fig.~\ref{fig:doctor} (b), demonstrate the clinical superiority of EndoControlMag. In the Easy Set evaluation, where six competing methods are assessed, EndoControlMag (Distance) achieves near-perfect average scores (9.75), significantly outperforming the second-best method, FlowMag~\cite{pan2024self} (average score 8.19), as well as other baseline approaches, which score considerably lower. This indicates a strong preference for our method under relatively ideal conditions. Crucially, the evaluation on the Hard Set confirms the robustness of EndoControlMag under challenging surgical conditions. Both variants of our method, i.e., Motion (9.19) and Distance (9.22), outperform FlowMag~\cite{pan2024self} (7.61) by a considerable margin.  The high scores awarded to our method, coupled with strong inter-rater agreement among the surgeons, underscore the effectiveness of our PRR and HTM frameworks in maintaining high-quality, artifact-free magnification despite occlusions, view changes, and tissue deformations. These findings robustly validate the clinical applicability and perceptual advantages of EndoControlMag for enhancing vascular visualization in complex surgical environments.

\begin{table*}[t]
\caption{\textbf{Ablation study investigating the impact of clip length $N$ in the Periodic Reference Resetting (PRR) mechanism.} Performance is evaluated on the Easy Set using image quality (SSIM, higher is better) and magnification accuracy ($E_{motion}$, lower is better) across five magnification factors ($\alpha \in \{2, 4, 8, 16, 32\}$) for varying clip lengths ($N \in \{2, 4, 6, 8, 10\}$). The results demonstrate that $N=4$ achieves the optimal balance between minimizing cumulative error and preserving temporal coherence, consistently yielding superior performance. Best results are highlighted in \textbf{bold}, with runner-up results \underline{underlined}.
\label{tab:abl_clip_length}}
\resizebox{\textwidth}{!}{
\begin{tabular}{lllllllllll}
\toprule
Metric     & \multicolumn{5}{c}{SSIM $\uparrow$}                                                                 & \multicolumn{5}{c}{$E_{motion}$ $\downarrow$}                                                              \\  \cmidrule(lr){2-6} \cmidrule(lr){7-11} 
Clip Length & $\times$2                   & $\times$4                   & $\times$8                   & $\times$16                  & $\times$32                  & $\times$2                   & $\times$4                   & $\times$8                   & $\times$16                   & $\times$32                    \\ \midrule
2           & 0.960$\pm$0.011          & 0.959$\pm$0.013          & 0.960$\pm$0.014          & 0.959$\pm$0.014          & 0.956$\pm$0.014          & \uline{1.001$\pm$0.558}         & 3.049$\pm$1.711          & 7.210$\pm$4.029          & \uline{15.069$\pm$8.703}          & {31.189$\pm$18.010}          \\
4           & \textbf{0.967$\pm$0.014} & \textbf{0.965$\pm$0.014} & \textbf{0.962$\pm$0.014} & \textbf{0.961$\pm$0.014} & \textbf{0.958$\pm$0.014} & \textbf{0.999$\pm$0.564} & \textbf{3.003$\pm$1.721} & \textbf{7.012$\pm$4.022} & \textbf{15.066$\pm$8.667} & \textbf{31.136$\pm$17.941} \\
6           & \textbf{0.967$\pm$0.014}          &\uline{0.964$\pm$0.014}           & 0.960$\pm$0.014          & \uline{0.960$\pm$0.014}          & \uline{0.957$\pm$0.015}          & 1.007$\pm$0.567          & \uline{3.018$\pm$1.721 }         & {7.670$\pm$3.593}          & 15.084$\pm$8.639          & \uline{31.155$\pm$17.917}          \\
8           & \uline{0.966$\pm$0.014}          & 0.963$\pm$0.014          & 0.960$\pm$0.014          & \uline{0.960$\pm$0.014}          & \uline{0.957$\pm$0.015}          & 1.018$\pm$0.571          & 3.037$\pm$1.721          & \uline{7.057$\pm$3.990}          & 15.114$\pm$8.615          & 31.179$\pm$17.895          \\
10          & 0.965$\pm$0.015          & {0.960$\pm$0.011}          & \uline{0.961$\pm$0.014}          & \uline{0.960$\pm$0.014}          & 0.956$\pm$0.015          & 1.027$\pm$0.577          & {3.446$\pm$1.970}          & {7.703$\pm$3.553}          & 15.123$\pm$8.597          & 31.194$\pm$17.882          \\ \bottomrule
\end{tabular}
}
\end{table*}

\subsection{Ablation Study}

\subsubsection{Optimizing PRR Clip Length}~\label{abl:N}
The clip length $N$ is a critical hyperparameter in our Periodic Reference Resetting (PRR) scheme, dictating the frequency of reference frame updates. This parameter represents an important trade-off: shorter clips (smaller $N$) minimize cumulative error but increase computational cost and may fragment continuous motion patterns, while longer clips (larger $N$) offer better computational efficiency but risk error accumulation, particularly during dynamic surgical events.
To determine the optimal balance, we conducted an ablation study evaluating the impact of varying clip lengths ($N \in \{2, 4, 6, 8, 10\}$) on magnification performance. We utilized the distance-based variant of EndoControlMag on the Easy Set and assessed performance using both image quality (SSIM) and magnification accuracy ($E_{motion}$) metrics across all five magnification factors ($\alpha \in \{2, 4, 8, 16, 32\}$).

The results, presented in Table~\ref{tab:abl_clip_length}, clearly indicate that a clip length of $N=4$ yields the best overall performance. This configuration consistently achieves the highest SSIM scores and the lowest $E_{motion}$ values across nearly all magnification factors. Performance degrades slightly with $N=2$, likely due to excessive reference frame switching disrupting the temporal continuity needed to capture smooth vascular pulsations. Conversely, longer clip lengths ($N \geq 6$) show a gradual decline in performance, confirming the detrimental effect of error accumulation over extended temporal windows.

\textbf{Physiological Relevance.}
The empirical superiority of $N=4$ aligns well with the physiological characteristics of surgical video data. Typical endoscopic systems operate at approximately 30 frames per second, while cardiac-induced vascular pulsations occur at 1-2 Hz (60-120 bpm). A clip length of $N=4$ corresponds to roughly 133 ms, allowing the PRR mechanism to reset the reference frame multiple times within a single pulsation cycle (which spans 15-30 frames). This frequency effectively bounds error accumulation while preserving sufficient temporal context to accurately represent the pulsatile motion. Based on this empirical evidence and physiological rationale, we adopt $N=4$ as the default clip length for EndoControlMag.

\begin{table*}[t]
\caption{\textbf{Ablation study of mask dilation strategies on the Easy Set.} We compare fixed-radius dilation with uniform weights and vessel-adaptive dilation with uniform weights or our adaptive weights with distance-based or motion-based softening. Performance is evaluated using SSIM (higher is better) and $E_{motion}$ (lower is better) across five magnification factors. Both adaptive softening strategies consistently outperform fixed-radius and uniform approaches, demonstrating the importance of anatomically and biomechanically informed mask design for artifact-free magnification. Best results are shown in \textbf{bold}, with runner-up results \underline{underlined}.}
\label{tab:dilation}
\resizebox{\textwidth}{!}{ 
\begin{tabular}{lllllllllllll}
\toprule
\multicolumn{3}{c}{Mask Dilation Strategy}              & \multicolumn{5}{c}{{SSIM $\uparrow$}}                                                                                                                              & \multicolumn{5}{c}{$E_{motion}$ $\downarrow$}                                                                                                                                \\ \cmidrule(lr){1-3} \cmidrule(lr){4-8} \cmidrule(lr){9-13} 
Dilation Radius $r$ & Soften (Distance) & Soften (Motion) & {$\times$2}          & {$\times$4}          & {$\times$8}          & {$\times$16}         & {$\times$32}         & {$\times$2}          & {$\times$4}          & {$\times$8}          & {$\times$16}          & {$\times$32}           \\ \cmidrule(lr){1-3} \cmidrule(lr){4-8} \cmidrule(lr){9-13}
Fixed 2.5 pixels             & \ding{55}             & \ding{55}           & {0.965$\pm$0.013} & {0.963$\pm$0.012} & {0.962$\pm$0.012} & {0.962$\pm$0.012} & {0.955$\pm$0.016} & {1.049$\pm$0.558} & {3.449$\pm$1.711} & {7.210$\pm$4.029} & {15.169$\pm$8.703} & {32.189$\pm$18.010} \\
Fixed 10 pixels             & \ding{55}             & \ding{55}           & {0.963$\pm$0.014} & {0.961$\pm$0.013} & {0.960$\pm$0.013} & {0.959$\pm$0.013} & {0.953$\pm$0.017} & {1.124$\pm$0.564} & {3.223$\pm$1.721} & {7.112$\pm$4.022} & {15.266$\pm$8.667} & {32.236$\pm$17.918} \\
Fixed 25 pixels             & \ding{55}             & \ding{55}           & {0.959$\pm$0.017} & {0.955$\pm$0.017} & {0.953$\pm$0.017} & {0.952$\pm$0.018} & {0.947$\pm$0.021} & {1.151$\pm$0.540} & {3.347$\pm$1.566} & {7.314$\pm$3.709} & {15.261$\pm$8.457} & {31.255$\pm$17.910} \\
Vessel-adaptive             & \ding{55}             & \ding{55}           & {0.961$\pm$0.015} & {0.965$\pm$0.011} & {0.960$\pm$0.013} & {0.960$\pm$0.017} & {0.953$\pm$0.016} & {1.087$\pm$0.904} & {3.290$\pm$1.892} & {7.197$\pm$4.009} & {15.152$\pm$8.394} & {31.203$\pm$17793}  \\
Vessel-adaptive
& \ding{52}             & \ding{55}         & {\uline{0.966$\pm$0.013}} & {\textbf{0.967$\pm$0.014}} & {\textbf{0.965$\pm$0.014}} & {\textbf{0.964$\pm$0.014}} & {\textbf{0.961$\pm$0.015}} & {\textbf{1.018$\pm$0.571}} & {\uline{3.037$\pm$1.721}} & {\textbf{7.057$\pm$3.990}} & {\textbf{15.114$\pm$8.615}} & {\textbf{31.179$\pm$17.895}} \\
Vessel-adaptive            & \ding{55}             & \ding{52}           & {\textbf{0.967$\pm$0.013}} & {\uline{0.966$\pm$0.014}} & {\uline{0.964$\pm$0.014}} & {\uline{0.964$\pm$0.010}} & {\uline{0.960$\pm$0.013}} & {\uline{1.027$\pm$0.577}} & {\textbf{3.003$\pm$1.510}} & {\uline{7.203$\pm$3.553}} & {\uline{15.123$\pm$8.597}} & {\uline{31.194$\pm$17.882}} \\ 
\bottomrule
\end{tabular}
}
\end{table*}

\subsubsection{Magnification Mask Dilation Strategy}
A critical component influencing the quality and realism of mask-conditioned magnification is the strategy used to define the transition zone surrounding the primary region of interest (ROI). This transition, achieved through mask dilation and subsequent weighting, dictates how magnification strength attenuates from the vessel core into the surrounding tissue. Conventional approaches, such as that used in FlowMag~\cite{pan2024self}, typically employ a fixed dilation radius and apply uniform magnification strength ($W_t=1$) within this dilated region. However, this simplistic approach fails to account for two crucial factors in surgical environments: the significant variation in vessel sizes across different anatomical locations and procedures, and the complex biomechanical interactions where vascular pulsations induce non-uniform, decaying displacements in adjacent tissues.

To address these limitations, our Hierarchical Tissue-aware Magnification (HTM) framework introduces two key designs for the outer mask region $M_t^{\text{out}}$: (1) \textit{vessel-adaptive dilation}, where the radius $r$ scales proportionally to the inner mask's dimensions ($r = \lfloor \gamma \cdot d_{\text{min}}(M_t^{\text{in}}) \rfloor$), ensuring the transition zone is appropriately sized relative to the vessel; and (2) \textit{spatially-varying softening}, which applies non-uniform magnification weights $W_t$ that gradually decrease from the vessel boundary outwards, mimicking natural biomechanical attenuation. We implement two distinct softening strategies: distance-based exponential decay (Eq.~\ref{eq:w_dist}) and motion-based weighting derived from optical flow (Eq.~\ref{eq:w_mot}).

To systematically evaluate the impact of these design choices, we conducted an ablation study comparing six distinct configurations on the Easy Set:
\begin{itemize}
    \item Fixed-radius dilation (2.5, 10, and 25 pixels) with uniform weights ($W_t=1$) – replicating FlowMag's strategy with varying radii.
    \item Vessel-adaptive radius with uniform weights ($W_t=1$) – isolating the effect of adaptive radius.
    \item Vessel-adaptive radius with distance-based softening – our first proposed HTM variant.
    \item Vessel-adaptive radius with motion-based softening – our second proposed HTM variant.
\end{itemize}

Table~\ref{tab:dilation} presents a comprehensive comparison across these configurations for both image quality (SSIM) and motion fidelity ($E_{motion}$). From the results, we can conclude that vessel-adaptive dilation consistently outperforms fixed-radius approaches across all magnification factors. This confirms our hypothesis that dilation should scale with vessel dimensions rather than using one-size-fits-all parameters.
When comparing our two softening strategies with vessel-adaptive dilation, distance-based softening achieves marginally better results in the Easy Set for higher magnification factors ($\times$8, $\times$16, $\times$32). This advantage likely stems from the stable, biomechanically-informed attenuation pattern of distance-based softening, which is particularly effective when vessels remain relatively stationary—a characteristic of the Easy Set.

\begin{figure}[t]
\centering
\includegraphics[width=\linewidth]{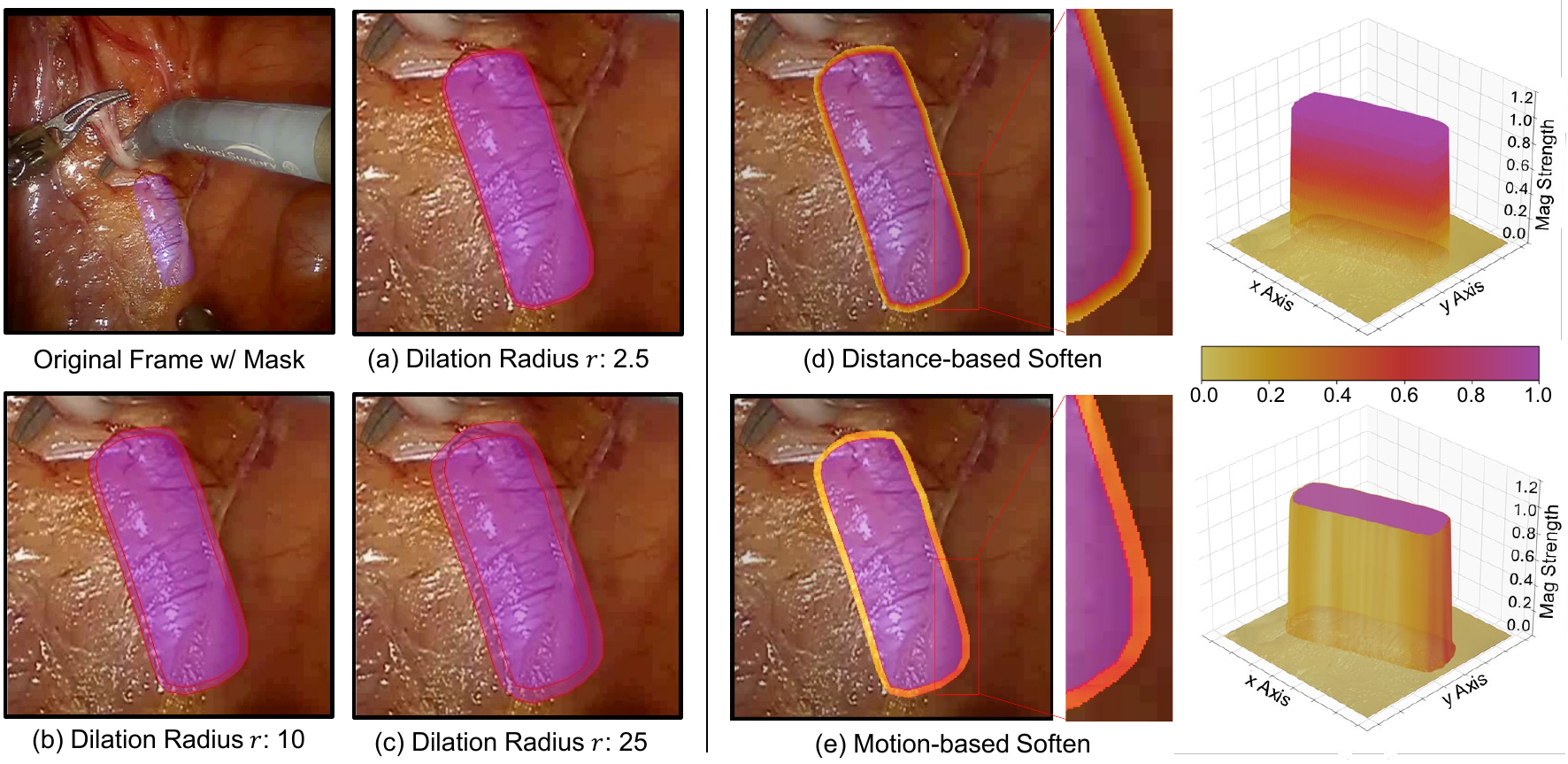}
\caption{\textbf{Visual comparison of mask dilation strategies for vascular motion magnification.} (a-c) Fixed-radius dilation with uniform weights (2.5, 10, and 25 pixels, respectively) creates abrupt boundaries between magnified and unmagnified regions. (d) Our distance-based softening generates a radially symmetric exponential decay pattern that simulates viscoelastic tissue attenuation. (e) Our motion-based softening produces a flow-guided pattern that adapts to measured tissue displacement. Both our adaptive methods (d, e) create biomechanically plausible transitions that respect tissue continuity and vascular-tissue mechanical coupling, whereas fixed-radius approaches (a-c) impose artificial boundaries that fail to model the graduated influence of vascular pulsations on surrounding tissues.}
\label{fig:dilation}
\end{figure}

Qualitative comparisons in Fig.~\ref{fig:dilation} visually corroborate these findings. Fixed-radius dilation with uniform weights (Figs.~\ref{fig:dilation}a-c) creates sharp, artificial boundaries between magnified and unmagnified regions, failing to model the gradual influence of pulsations on surrounding tissue. In contrast, our distance-based softening (Fig.~\ref{fig:dilation}d) generates a smooth, radially decaying gradient, resembling a heat map that aligns well with the expected biomechanical attenuation of forces in elastic media. Motion-based softening (Fig.~\ref{fig:dilation}e) produces a more complex pattern reflecting the actual measured tissue displacements, capturing potentially asymmetric responses influenced by local tissue properties and constraints.

The superior performance of our adaptive approaches reflects the fundamental biomechanical properties of vascular-tissue interactions. Traditional fixed-radius approaches implicitly assume uniform tissue elasticity throughout the surgical field, contradicting the heterogeneous nature of biological tissues. Our vessel-adaptive strategies recognize that larger vessels typically influence a proportionally larger surrounding area due to greater pulsation amplitude and tissue displacement.
Furthermore, both softening strategies model the elastic coupling between vessels and surrounding tissues from different but complementary perspectives. Distance-based softening reflects the natural attenuation of mechanical waves in viscoelastic media, while motion-based softening directly captures the empirical displacement patterns resulting from these physical interactions.

The results validate our HTM framework, demonstrating that both adaptive radius and adaptive softening are essential for achieving high-fidelity, artifact-free magnification. While distance-based softening shows a slight edge in the stable Easy Set, the complementary nature of the two strategies provides flexibility for handling the more complex dynamics encountered in the Hard Set, as discussed previously.

\section{Conclusion and Discussion}

This paper introduces EndoControlMag, a robust framework for vascular motion magnification in endoscopic surgery that addresses key challenges in surgical vision enhancement. Our training-free Lagrangian approach with hierarchical mask-conditioned control introduces two complementary designs: Periodic Reference Resetting (PRR), which prevents error accumulation by dynamically updating reference frames, and Hierarchical Tissue-aware Magnification (HTM), which enables biomechanically-informed spatial control through adaptive vessel tracking and softening strategies.

Comprehensive evaluation across diverse surgical scenarios on our EndoVMM24 dataset demonstrates that EndoControlMag consistently outperforms existing methods in both image quality and magnification accuracy. Our framework exhibits particular robustness in challenging conditions frequently encountered in clinical practice, including occlusions, view changes, tissue deformations, and tool disturbance. The two softening variants offer complementary advantages: motion-based softening excels with complex tissue deformations, while distance-based softening provides stability when optical flow estimation becomes unreliable.

\subsection{Clinical Implications}

The ability to selectively enhance subtle vascular pulsations while preserving surrounding tissue integrity has significant clinical implications. By providing surgeons with enhanced visualization of blood vessels, EndoControlMag could potentially reduce the risk of vascular injuries during dissection and resection procedures, particularly in anatomically complex regions. The improved identification of critical vascular structures enables more precise surgical navigation and tissue manipulation. Furthermore, the enhanced visualization reveals physiological information through pulsation patterns, which may inform surgical decision-making regarding tissue viability and perfusion status. Less experienced surgeons, especially, may benefit from these magnified visual cues, which make subtle anatomical details more apparent and potentially accelerate the learning curve for complex minimally invasive procedures. During lengthy operations, the technology could reduce cognitive load by highlighting key anatomical features, allowing surgeons to maintain focus on critical structures throughout the procedure. The interactive nature of our framework, which allows surgeons to designate regions of interest and adjust magnification strength, aligns well with the surgeon-in-the-loop paradigm essential for clinical adoption of AI-augmented visualization technologies.

\subsection{Limitations and Future Work}
Despite promising results, several limitations warrant discussion. The robustness of our framework partially relies on the performance of pre-trained, off-the-shelf models for optical flow (RAFT~\cite{teed2020raft}) and tracking (MFT~\cite{neoral2024mft}). These models were utilized directly without domain-specific fine-tuning on surgical data. Consequently, their performance may degrade under conditions significantly different from their original training distributions, potentially leading to failure modes. For instance, the VOT tracker may lose the target vessel during extreme occlusions (e.g., $>$75\% vessel area obscured by dense smoke or instruments for extended periods, $>$2 seconds) or rapid camera movements (e.g., $>$100 pixels/frame displacement causing significant motion blur or the vessel exiting the field of view entirely). Such tracker failures would necessitate manual reinitialization by the user selecting the vessel mask again, interrupting the workflow. Similarly, optical flow estimation, crucial for the motion-based softening strategy, can become unreliable under poor illumination, heavy smoke, specular reflections, or extremely fast, non-rigid tissue deformations, potentially degrading the quality of motion-based softening. While our PRR scheme mitigates long-term drift, very abrupt motions within a single short clip ($N=4$) could still momentarily challenge flow estimation accuracy.

Furthermore, the selection between the motion-based and distance-based softening strategies currently requires manual input or pre-selection based on the anticipated surgical context. This allows surgeons to prioritize motion-based softening when tissue deformation is prominent and optical flow is reliable, or switch to the more stable distance-based softening during periods of heavy smoke or instrument occlusion where flow estimation is compromised. However, this manual switching adds a layer of user interaction. An automated, context-aware mechanism that dynamically selects the optimal softening strategy based on real-time assessment of scene conditions (e.g., smoke presence, flow quality metrics) could significantly enhance clinical utility and represents an important direction for future work.

Regarding computational performance, the enhanced robustness and adaptability of EndoControlMag, particularly the integration of the PRR scheme, recursive mask tracking (VOT), and adaptive softening calculations, introduce additional processing overhead compared to simpler baseline methods like FlowMag~\cite{pan2024self}. On our test hardware (NVIDIA RTX A6000), processing a single frame takes approximately 2 seconds in total. This processing time currently limits the applicability for seamless real-time integration into live surgical video feeds, which typically demand frame rates exceeding 25-30 FPS. However, the current performance is adequate for offline applications, such as post-operative surgical review or analysis, and may be suitable for guidance in specific scenarios where immediate feedback is not paramount. Achieving real-time performance will necessitate further optimization, potentially through model distillation, dedicated hardware acceleration, exploring lighter-weight alternatives for optical flow and tracking components, or optimizing the implementation for surgical system integration.

Future research should focus on addressing these limitations. Domain adaptation or targeted fine-tuning of the pre-trained flow and tracking models using surgical data could improve their robustness to specific intraoperative challenges. Developing more sophisticated tracking algorithms resilient to long-term occlusions and appearance changes is crucial. Integrating depth information from stereoscopic endoscopes could enable depth-aware magnification, improving specificity in complex 3D anatomies. 
Development of quantitative metrics derived from magnified pulsations could support clinical decision-making through extracted hemodynamic parameters. Finally, comprehensive multi-specialty user studies remain essential for validating clinical utility and refining workflow integration.

In conclusion, EndoControlMag represents a promising solution in surgical vision enhancement by providing robust, interactive, and contextually aware vascular motion magnification. By addressing the unique challenges of endoscopic environments while maintaining high visual fidelity, our approach has the potential to improve surgical precision and patient outcomes across a wide range of minimally invasive procedures.



 \bibliographystyle{elsarticle-harv} 
 \bibliography{main}



\end{document}